\documentclass[aps,prd,amsmath,amssymb,showkeys,showpacs,10pt]{revtex4}
\usepackage{amsfonts}
\usepackage{graphicx}
\usepackage{bm}
\def\journal#1#2#3#4{{#1} {\bf #2}, #3 (#4)}

\newcommand{\be}{\begin{equation}}
\newcommand{\ee}{\end{equation}}
\newcommand{\bea}{\begin{eqnarray}}
\newcommand{\eea}{\end{eqnarray}}
\newcommand{\hf}{\frac12}
\newcommand{\nn}{\nonumber\\}
\def\eq#1{(\ref{#1})}
\def\la{\langle}
\def\ra{\rangle}
\def\Tr{{\mathrm{Tr}}}
\def\tr{{\mathrm{tr}}}
\def\mr#1{{\mathrm{#1}}}
\def\v#1{{\bm{#1}}}
\def\ord#1{{\cal O}(#1)}
\def\br{\hskip-6pt/}

\def\fd#1#2{\frac{\delta#1}{\delta#2}}

\def\etab{\bar\eta}
\def\hamma{\hat\gamma}
\def\heta{\hat\eta}
\def\hetab{\hat{\bar\eta}}
\def\hj{\hat j}
\def\hpsi{\hat\psi}
\def\hpsib{\hat{\bar\psi}}
\def\hA{\hat A}
\def\hD{\hat D}
\def\hG{\hat G}

\def\kfer{k}
\def\psib{\bar\psi}
\def\ub{\bar u}

\begin{document}
\title{Scattering in an environment}
\author{Janos Polonyi$^a$, Karima Zazoua$^{a,b}$}
\affiliation{a: Strasbourg University, High Energy Theory Group, CNRS-IPHC,
23 rue du Loess, BP28 67037 Strasbourg Cedex 2 France}
\affiliation{b: LEPM USTO-MB, BP 1505 EL M`naouer Oran, Algeria}
\date{\today}

\begin{abstract}
The cross section of elastic electron-proton scattering taking place in an electron gas is calculated within the Closed Time Path method by means of the resummation of the one-loop self-energy in the photon propagator. Back-reaction of the colliding particles on the gas has been taken into account and is found to be dominating the cross section when the energy exchange falls in the vicinity of the Fermi energy. Back-reaction reflects the colliding particles-gas entanglement and makes the colliding particle state mixed. The softness of the asymptotic particle-hole states of the gas makes the colliding particle trajectories more consistent and the collision irreversible, rendering in this manner the scattering more classical in this regime.
\end{abstract}
\maketitle

\section{Introduction}
Scattering experiments are imagined as idealized processes, described by transition probabilities between asymptotic in- and out-states without interactions, interpolated in time by turning on and off adiabatically interactions. In reality these processes take place in an environment that interacts with the system even if the final state interactions are neglected. The influence of the environment on the asymptotic states can be taken into account in a simple manner as long as the environment possesses no asymptotic states. For instance, virtual vacuum polarizations might be considered as environment from the point of view of the colliding particles and their influence on the final state can be accounted by using dressed, quasifree particles in the asymptotic states. The influence of the environment on the cllision process is then restricted to the collision zone and can easily be recovered by the usual perturbation expansion.

This problem becomes more interesting when the environment has asymptotic states. The difference between system and environment is that the former evolves from a fixed initial state to a given final one, the latter has initial conditions only: it is not identified by the detectors hence it follows an open-ended, undisturbed time evolution. One can, in principle, take such a possibility into account within the usual description of scattering, by summing the transition probability over possible orthogonal environment final states. But, this is not feasible in practice where this summation is carried out explicitly due to the large number of possible final environment states. A way out from this problem is to build in the trace over environment states into the transition probability form the very beginning. The closed time path (CTP) method proposed by  J. Schwinger \cite{schw} proves to be the appropriate framework for this problem. It is not only tailored to calculate expectation value but transition probability can be recast in this form. In addition, it contains the trace we need, over the Fock space of the theory to arrive at probabilities instead of transition amplitudes. An additional bonus of using this scheme is a simple way to build up the entanglement between the colliding particles and their environment and the mixing in the reduced density matrix for the scattering particles. 

The goal of this  paper is the description of the electron-proton elastic scattering taking place in a homogeneous electron gas. We consider the gas at vanishing temperature for the sake of simplicity; therefore, the model starts with a pure state, a Fermi-sphere of electrons, no photons, and an incident proton, at the initial time. The final state is identified by the momentum of the proton and a recoiled electron, being originally at rest with respect to the gas. This problem may be considered as the first step toward a more realistic description of collision processes where both the beam and the target are systems consisting of a large number of particles. Our main result is the form of the cross section as a sum of the expected expression containing the photon propagator dressed by the environment and another term, comprising the effects of colliding particles-environment entanglement. There are kinematical regions where this latter is more important than the former. We monitor the consistency \cite{griffithe,griffithk,dowker} of the colliding particle trajectories and the dynamical breakdown of time reversal invariance, two necessary conditions for classical behavior. We find that both of them are strong for scattering where the entanglement contribution is dominant, supporting the expectation about the importance of entanglement in classical limit.

The CTP method has already been used in different forms with different goals. It emerges in a thorough analysis of many-body theory \cite{kadanoff}; it was introduced in solid state physics \cite{keldysh} to calculate retarded Green functions. The discussion of quantum field theory at finite temperature and/or finite density is based on this formalism \cite{niemisa,niemisb,niemi} and algebraically defined thermal field theory (TFT) \cite{umezawa} can be derived from it, as well. Its path integral setting \cite{kobes} opened the way for a large number of applications \cite{calzettab}. Relaxation has been studied in an extensive manner \cite{berges} by deriving quantum kinetic equations in two different ways. One is to use the Schwinger-Dyson, alias Kadanoff-Baym equations \cite{kadanoff} to arrive at a Boltzmann-equation-like balance condition in a plasma  \cite{li,calzettahu,mrowczynski,klevansky,koike,prokopec,anisimov}. Another way is to construct the analogy of a quantum Brownian motion \cite{caldeira}, a Langevin-like equation in field theory \cite{chou,yokoyama,boyanovsky}. 

The problem addressed here: to follow a single collision in an environment is a nonequilibrium process and requires the use of the CTP formalism that is equivalent with a special, unusual realization of TFT where one has access to transition probabilities and can trace the effects of entanglement between the colliding particles and their environment and estimate the consistency of the collision. The cross section is obtained by relying on the usual a simplification, adiabatic switching. It is assumed that interactions are weak when the proton enters into the electron gas and when it leaves together with the kicked electron. Apart from some initial steps in TFT \cite{manoukian,varma,hong}, we are not aware of the systematic application of the CTP scheme to describe transition probabilities or scattering cross sections. Another novel aspect of the results presented below is that the back-reaction of the collision process on the environment, a rather complicated effect that is usually ignored in the quantum kinetic approach, is retained. It can be taken into account in a simple manner by means of an explicit, algebraic solution of the Kadanoff-Baym equation. The resulting full-fledged interaction with balanced action-reaction forces leads to colliding charge-gas entanglement. Such an entanglement is incorporated in the photon self-energy, representing the coupling of photon to particle-hole excitations. The entanglement contributions dominate the cross section in the kinematical regime where the asymptotic excitations of the electron gas have sufficiently soft excitation spectrum. This softness makes entanglement nonperturbative and suppresses the quantum interference of trajectories thereby driving the collision classical.

We start in Section \ref{collgas} with the outline of an idealized version of an electron-proton scattering in an electron gas. The way of recognizing entanglement, mixing in the reduced density matrix and consistency is explained in Section \ref{entangl}. The reduction formula and its perturbative evaluation is discussed in Section \ref{transition}. The partially resummed photon propagator, appearing in the expression for the transition probability and its spectral functions are obtained in Section \ref{photprop}. The numerical results for the cross section are presented in Section \ref{scattres}. Finally, the summary of our findings is given in Section \ref{summary}.

\section{Collision in a gas}\label{collgas}
We consider in this work elastic electron-proton scattering taking place in a homogeneous electron gas at vanishing temperature and finite density. A proton falls on the gas and electrons are knocked out from the gas by the collision. Though the intial state of electrons participating in the collision is not controllable we assume that those scattering events are kept where the initial electron was at rest with respect to the gas. 

The adiabatic switching hypothesis is more involved in this case than for scatterings in the vacuum because of the interacting environment. Let us distinguish the following well-separated times: $t_i\ll t_{in}\ll t_{out}\ll t_{final}$. First a noninteracting electron and proton beam are created at the initial time $t_i$. The protons enter into the electron gas at $t_{in}$, the colliding particles leave the gas at $t_{out}$, and finally, the assembly of noninteracting final states of the colliding particles and the gas is removed at the final time $t_{final}$. The adiabatic switching of the electromagnetic interaction should satisfy two constraints. On the one hand, the interactions should be strong enough to build up or deconstruct the interacting electron gas before $t_{in}$ or after $t_{out}$, respectively. On the other hand, the interactions should be weak enough at $t_{in}$ and $t_{out}$ to render the entering into and leaving from the electron gas adiabatic, leaving behind no excitations.

The initial state $|\Psi_i\ra=b_{p_i}^\dagger a_{e_i}^\dagger|\Psi^{e_i}_{gas}\ra$ is written as a product of the operators $b_{p_i}^\dagger$ and $a_{e_i}^\dagger$, which create the initial proton and electron states, respectively and
\be
|\Psi^{e_i}_{gas}\ra=\prod_{e_i\ne j\in gas}a_j^\dagger|0\ra,
\ee
denotes the state of the free gas with a hole at the quantum number $e_i$. The amplitude of transition to a state containing a free proton and electron with quantum numbers
$p_f$ and $e_f$ is
\be
{\cal A}_n=\la n|a_{e_f}b_{p_f}U(t_f,t_i)|\Psi_i\ra.
\ee
where $n$ denotes the quantum number of the remaining electron-photon component of the final state and $U(t_f,t_i)$ stands for the time evolution operator. It is assumed that the detectors register the state of the colliding particles only and the electrons in the rest of the gas and photons are left in an arbitrary final state. The probability of the observed transition is therefore the sum over possible states of the remainder of the electron gas and photons, 
\be\label{trprobop}
P(f\leftarrow i)=\sum_n|{\cal A}_n|^2,
\ee
which can be rewritten as
\be\label{trprobsum}
P(f\leftarrow i)=\sum_n\la\Psi_i|U^\dagger(t_f,t_i)b_{p_f}^\dagger a_{e_f}^\dagger|n\ra\la n|a_{ef}b_{p_f}U(t_f,t_i)|\Psi_i\ra.
\ee

Baryon and lepton number conservations allow us to represent the summation over all electron and photon basis state compatible with the initial conditions by the insertion of identity,
\be\label{trproba}
P(f\leftarrow i)=\la\Psi_i|U^\dagger(t_f,t_i)b_{p_f}^\dagger a_{e_f}^\dagger a_{ef}b_{p_f}U(t_f,t_i)|\Psi_i\ra.
\ee
This expression is finally written as
\be\label{trprob}
P(f\leftarrow i)=\Tr[{\cal O}U(t_f,t_i)\rho_iU^\dagger(t_f,t_i)]
\ee
where the trace is taken over the full Fock space, 
\be
\rho_i=b_{p_i}^\dagger a_{e_i}^\dagger|\Psi^{e_i}_{gas}\ra\la\Psi^{e_i}_{gas}|a_{e_i}b_{p_i}
\ee
denotes the initial density matrix, and the Hermitean operator
\be\label{trprobobs}
{\cal O}=b_{p_f}^\dagger a_{e_f}^\dagger a_{e_f}b_{p_f}
\ee
handles the final state of the colliding particles. It is a well known procedure to sum transition probability over unresolved final states. But, the sum in Eq. \eq{trprobsum} is over a too-large space and cannot be handled in the usual, explicit manner. The advantage of the form \eq{trprob} is that it is an expectation value of a single observable incorporating the summation over the final states in Eq. \eq{trprobop}.

\section{System-environment entanglement}\label{entangl}
Before applying perturbation for the expectation value \eq{trprob}, we point out the advantage of this formalism in finding the influence of an environment on the system studied. System stands for the collections of degrees of freedom with time evolution followed, environment denotes the remaining part of dynamics. The boundary conditions in time differ for these two components: the environment always follows open-ended, free time evolution encoded as an initial condition problem. The system may obey both intial and final conditions, like in scattering experiments. 

Let us introduce the generic field variables $\phi$ and $\chi$ for our system and its environment and the action $S[\phi,\chi]$ governing the dynamics. The initial state is given by the density matrix $\rho_i[(\tilde\phi^+,\tilde\chi^+),(\tilde\phi^-,\tilde\chi^-)]$ where the fields with tilde depend on spatial coordinates only, $\tilde\phi=\tilde\phi(\v{x})$, etc.. Our goal is to find the reduced density matrix
\be\label{reddensmint}
\rho_s[\tilde\phi^+,\tilde\phi^-]=\int D[\tilde\chi]\rho[(\tilde\phi^+,\tilde\chi),(\tilde\phi^-,\tilde\chi)],
\ee
for the system.

Consider first the complete density matrix subject of the time evolution
\be\label{tddensm}
\rho_t=U(t,t_i)\rho_iU^\dagger(t,t_i).
\ee
The reduced density matrix is obtained by tracing out the environment suggesting the introduction of the generator functional
\be\label{reddenspi}
e^{iW[\hj;\tilde\phi^\pm]}=\la\tilde\phi^+|\Tr_e\bigl[ T[e^{-i\int_{t_i}^{t_f}dx^0\int d^3x[H(x)-j^+(x)\phi(x)]}]\rho_iT^*[e^{i\int_{t_i}^{t_f}dx^0\int d^3x[H(x)+j^-(x)\phi(x)]}]\bigr]|\tilde\phi^-\ra
\ee
where the trace is over the environment and $T^*$ denotes antitime ordering. The path integral representation of this functional,
\be\label{otppi}
e^{iW[\hj;\tilde\phi^\pm]}=\int D[\hat\phi]D[\hat\chi]e^{iS[\phi^+,\chi^+]-iS[\phi^-,\chi^-]+i\hj\cdot\hat\phi}.
\ee
The boundary conditions in time, suppressed in this equation for the sake of better readability are the following. There is an integration over the initial field configurations with the weight factor $\rho_i[(\tilde\phi^+,\tilde\chi^+),(\tilde\phi^-,\tilde\chi^-)]$. At the final time we impose CTP boundary conditions for the environment, $\chi^+(t_f,\v{x})=\chi^-(t_f,\v{x})$ to assure an unconstrained time evolution for the system, the integration of the final configuration $\tilde\chi$ corresponding to the trace
operation in the environment sector in Eqs. \eq{reddensmint}, \eq{reddenspi}. Finally, the system trajectories obey open time path (OTP) boundary conditions, $\phi^\pm(t_f,\v{x})=\phi^\pm_f(\v{x})$. 

We write now the bare action of the theory as the sum of a system action and the rest, $S[\phi,\chi]=S_s[\phi]+S_e[\phi,\chi]$, and consider the effective theory for the system by integrating over the environment variables,
\be\label{reddensm}
e^{iW[\hj;\tilde\phi^\pm]}=\int D[\hat\phi]e^{iS_s[\phi^+]-iS_s[\phi^-]-iS_I[\phi^+,\phi^-]+i\hj\cdot\hat\phi}
\ee
where the influence functional $S_I[\phi^+,\phi^-]$ \cite{feynman} is defined as
\be\label{inflfunc}
e^{iS_I[\phi^+,\phi^-]}=\int D[\hat\chi]e^{iS_e[\phi^+,\chi^+]-iS_e[\phi^-,\chi^-]}.
\ee
The effective vertices, terms in the influence functional, can be classified as direct and entangled, the latter coupling $\phi^+$ and $\phi^-$, $S_I[\phi^+,\phi^-]=S_d[\phi^+]-S_d[\phi^-]+S_e[\phi^+,\phi^-]$, $\delta^2S_e/\delta\phi^+\delta\phi^-\ne0$.
In the OTP scheme interactions are represented by couplings among fields on the same time axis, eg. the interactions within the system give rise $S_d[\phi^\pm]$. Each contribution of the path integral \eq{otppi} for a given pair of environment configurations $\chi^\pm$ gives a factorisable contribution to the density matrix, $\rho_i[\tilde\phi^+,\tilde\phi^-;\chi^+,\chi^-]$ which represents a pure system state. The CTP boundary conditions for the environment couple $\chi^+$ with $\chi^-$ and generate $S_e[\phi^+,\phi^-]$. It is the integration over the environment configurations in Eq. \eq{inflfunc} that produces a density matrix of a mixed state. 

The path integral representation of the reduced density matrix makes two phenomena related to the classical limit, consistency and decoherence, particularly explicit. The smallness of the off-diagonal matrix elements, the linear superposition of different coordinate $\tilde\phi$ states, is a measure of decoherence \cite{zeh}, the quality of $\tilde\phi$ as a pointer variable \cite{zurek}. Decoherence is induced by the interactions with  the environment. Let us now consider a pair of system trajectories $\phi^\pm$, their contribution to the reduced density matrix is the integrands on the right-hand side of Eq. \eq{reddensm}, which is suppressed due to interactions with the environment when $\Im S_I[\phi^+,\phi^-]\ll0$ ($\hbar=1$). Hence, the more negative is the imaginary part of the influence functional, the more decohered is the pair $\phi^\pm$.

Decoherence characterizes the reduced density matrix at a given instant of time. It builds up in time and this dynamical process is called consistency \cite{griffithe,griffithk,dowker}. Let us consider now two trajectories $\phi^+$ and $\phi^-$, $\phi^+(t_f,\v{x})=\phi^-(t_f,\v{x})=\tilde\phi(\v{x})$ contributing to the diagonal matrix element in Eq. \eq{reddensm} at $\tilde\phi$. There are four contributions, four pairs or trajectories related to $\phi^+$ and $\phi^-$, namely $\hat\phi=(\phi^+,\phi^+)$,  $(\phi^+,\phi^-)$,  $(\phi^-,\phi^+)$ and $(\phi^-,\phi^-)$. This pair of trajectory is called consistent if their contribution to the probability of finding $\tilde\phi$ in the final state is additive. This situation is  approached when the contribution of $\hat\phi=(\phi^+,\phi^-)$ is small, $\Im S_I[\phi^+,\phi^-]\ll0$.

Relation between decoherence and consistency is clear: A pair of trajectories is consistent if the imaginary part of the influence functional of the CTP formalism receives sufficiently negative contributions from the time $t'$ of propagation when these trajectories are different, $\phi^+(t',\v{x})\ne\phi^-(t',\v{x})$. But stopping the time evolution at such a time and considering the contribution in OTP scheme this integrand represents a contribution to the reduced density matrix $\rho_s[\phi^+(t',\v{x}),\phi^-(t',\v{x})]$ and indicates decoherence at that particular time. In other words, decoherence sustained in sufficiently long time makes consistency.

\section{Transition probability}\label{transition}
We introduce now the reduction formula for the electron-proton elastic scattering, together with its perturbative evaluation.

\subsection{Reduction formulas}
The transformation of the transition probability of an inclusive scattering process, \eq{trprobsum} into an expectation value \eq{trprob} requires the use of the reduction formulas \cite{lehman}. It is obvious to generalize this scheme for the CTP formalism and the result can be easily presented in terms of the generator functional for connected Green functions, constructed by introducing external sources coupled linearly to the elementary fields in the Lagrangian. To handle both time evolution operators in Eq. \eq{trprob} one introduces independent external sources for $U$ and $U^\dagger$, indexed with + and -, respectively, $\heta_e=(\eta_e^+,\eta_e^-)$ and $\heta_p=(\eta_p^+,\eta_p^-)$ for the electron and proton fields and $\hj=(j^+,j^-)$ for the gauge field. The generator functional is, therefore, defined by
\bea\label{waj}
e^{iW[\hj,\hetab_\tau,\heta_\tau]}&=&\Tr\bigl[U(t_f,t_i;\eta^+,\etab^+,j^+)\rho_iU^\dagger(t_f,t_i;-\eta_-,-\etab_-,-j^-)\bigr]
\eea
($\hbar=c=1$) where 
\be
U(t_f,t_i;\eta,\etab,j)=T[e^{-i\int_{t_i}^{t_f}dx^0\int d^3x[H(x)-\sum_\tau(\etab_\tau(x)\psi_\tau(x)+\psib_\tau(x)\eta_\tau(x))-j^\mu(x)A_\mu(x)]}]
\ee
denotes the time evolution operator in the presence of the external sources $\eta,\etab,j$, and $H(x)$ stands for the energy density, $\tau=e$ or $p$. Each degree of freedom appears twice, once in the time ordered exponentialized form of $U(t_i,t_f)$ and another one in its Hermitian conjugate, involving antitime ordering $T^*$. The path integral representation of \eq{waj} is
\be\label{wpint}
e^{iW[\hj,\hetab_\tau,\heta_\tau]}=\int\prod_\tau D[\hpsi_\tau]D[\hpsib_\tau]D[\hA]e^{iS[\hA,\hpsib_\tau,\hpsib_\tau]+i\hetab_\tau(x)\hpsi_\tau(x)+i\hpsib_\tau(x)\heta_\tau(x)]+i\int dx\hj(x)\hA(x)}
\ee
where the integration is over pairs of trajectories $\hpsi_\tau=(\psi_\tau^+,\psi_\tau^-)$, $\hA=(A^+,A^-)$ and the action
\be\label{ctpact}
S[\hA,\hpsi_\tau,\hpsib_\tau]=\sum_\tau\int dx\hpsib_\tau(x)(\hG_\tau^{-1}-e_\tau\hA\br(x))\hpsi_\tau(x)+\hf\int dx\hA(x)\hD_0^{-1}\hA(x)
\ee
contains $\hA\br=A_\mu\hamma^\mu$ with the Dirac doublet matrices $\hamma^\mu=(\gamma^\mu,-(\gamma^\mu)^*)$. The fermion inverse propagators are
\be
\hG_\tau^{-1}=\begin{pmatrix}i\partial\br-m_\tau+i\epsilon&0\cr0&-\gamma^0(i\partial\br
m_\tau+i\epsilon)^\dagger\gamma^0\end{pmatrix}+\hG_{BC\tau}^{-1}
\ee
$m_e=m$ and $m_p=M$ being the electron and proton mass, respectively. The inverse photon propagator
\bea
\hD^{-1}_{0\mu\nu}&=&g_{\mu\nu}\begin{pmatrix}\Box+i\epsilon&0\cr0&-\Box+i\epsilon\end{pmatrix}+\hD^{-1}_{BC},
\eea
is given in Feynman gauge. The boundary conditions in time are
\bea\label{bcemf}
\psi_\tau^+(t_f,\v{x})&=&\psi_\tau^-(t_f,\v{x}),\nn
\psib_\tau^+(t_f,\v{x})&=&\psib_\tau^-(t_f,\v{x}),\nn
A^+_\mu(t_f,\v{x})&=&A^-_\mu(t_f,\v{x}),
\eea
due to the trace in Eq. \eq{waj}. The closing of the path at the final time can be implemented on the level of the free propagators. Since the free propagators can be most easily derived in the operator formalism the actual form of $\hG^{-1}_{BC}$ and $\hD^{-1}_{BC}$ will not be important for us. 

One has to introduce a gauge invariant cutoff, for instance, dimensional regularization 
after Wick rotation, and the corresponding counterterms in the exponent of Eq. \eq{wpint}.
Since the boundary conditions in time do not affect, the UV divergences the renormalization
and the counterterm structure are identical with the single time axis formalism and will be
suppressed in what follows. To remove IR divergent tadpole contributions we assume the presence of a classical, homogeneous external charge density that neutralizes the electron gas. 

We can finally turn to the transition probability \eq{trprob}, which can be obtained by means of the generator functional introduced above by repeating the steps followed in deriving the reduction formulas. The transition probability of the elastic scattering process $e(p_1)+p(p_2)\to e(q_1)+p(q_2)$ reads as
\bea\label{trprobfd}
P(f\leftarrow i)&=&Z_e^{-2}Z_p^{-2}\int dx^+_1dx^+_2dy^+_1dy^+_2dx^-_1dx^-_2dy^-_1dy^-_2
e^{iq_1(y^+_1-y^-_1)+iq_2(y^+_2-y^-_2)-ip_1(x^+_1-x^-_1)-ip_2(x^+_2-x^-_2)}\nn
&&\times[\ub_1(i\partial\br_{y^+_1}-m)]_{\beta^+_1}[\ub_2(i\partial\br_{y^+_2}-M)]_{\beta^+_2}[\ub^*_1(-i\partial\br_{y^-_1}-m)]_{\beta^-_1}[\ub^*_2(-i\partial\br_{y^-_2}-M)]_{\beta^-_2}\nn
&&\times[(-i\partial\br_{x^+_1}-m)u_1]_{\alpha^+_1}[(-i\partial\br_{x^+_2}-M)u_2]_{\alpha^+_2}[(i\partial\br_{x^-_1}-m)u^*_1]_{\alpha^-_1}[(i\partial\br_{x^-_2}-M)u^*_2]_{\alpha^-_2}\nn
&&\times\frac{\delta^8W[\hj,\hetab,\heta]}{\delta\eta_{p\alpha^-_2}^-(x^-_2)\delta\eta_{e\alpha^-_1}^-(x^-_1)\delta\etab_{p\beta^-_2}^-(y^-_2)\delta\etab_{e\beta^-_1}^-(y^-_1)\delta\etab_{e\beta^+_1}^+(y^+_1)\delta\etab_{p\beta^+_2}^+(y^+_2)
\delta\eta_{e\alpha^+_1}^+(x^+_1)\delta\eta_{p\alpha^+_2}^+(x^+_2)}_{|\hj=\heta_\tau=\hetab_\tau=0}.
\eea
where $Z$ denotes the wave function renormalization constant, expressing the proportionality of asymptotic and interpolating fields, $\psi_e=Z_e^{1/2}\psi_e^{as}$ and $\psi_p=Z_p^{1/2}\psi_p^{as}$. The Dirac spinors $u_j$ and $\ub_j$ describe the initial and final states, respectively, for electrons ($j=1$) and protons ($j=2$). This expression will be evaluated below by a partial resummation of the perturbation series.

The calculation of scattering probability brings out an important difference between the TFT and the CTP formalism. In TFT, the forward-pointing time axis is used to follow physical processes and the other time axis that contains the so-called ghost degrees of freedom is a formal device to calculate the thermal averages with real time dependence. The corresponding generator functional for QED is 
\be\label{tftgenf}
Z_{TFT}=\Tr[U(t_f,t_i;\eta^+,\etab^+,j^+)\rho^{1-s}_iU^\dagger(t_f,t_i;-\eta^-,-\etab^-,-j^-)\rho^s_i],
\ee
where $\rho_i$ is the density matrix in thermal equilibrium and $0\le s\le 1$. It is easy to see that the expectation values obtained by functional derivatives with respect to the external sources $\eta^+$, $\etab^+$ and $j^+$ evaluated at $\heta=\hetab=\hj=0$ are independent of $s$, whose value can be freely set to simplify the free Green functions. One can obtain in this manner the thermal average of an observable or the transition probability of a process as long as equilibrium prevails, namely the condition $[\rho^s_i,U(t,t';0)]=0$ holds after carrying out the observation or completing the process. But, it may happen that the measurement or the process generates nonequilibrium phenomenon such as the decay of the ``vacuum polarization cloud'' in the gas, made by the colliding particles. In such a case, we need $s=0$, the choice that makes the functionals \eq{waj} and \eq{tftgenf} identical, renders the ghost degrees of freedom physical and brings TFT equivalent with CTP. Another special feature of CTP, the choice $s=0$, is that it makes the reduced density matrix available and opens the way to address the issue of quantum-classical transition in expectation values or in processes.

\subsection{Perturbation expansion}
The free generating functional,
\be
W_0[\hj,\hetab_e,\heta_e,\hetab_p,\heta_p]=-\int dxdy\left[\hf\hj(x)\hD_0(x,y)\hj(y)+\sum_\tau\hetab_\tau(x)\hG_{0\tau}(x,y)\heta_\tau(y)\right]
\ee
contains the CTP propagators with Fourier transforms
\bea
\hG_{0\tau}(q)&=&\int dx\hG_{0\tau}(x,0)e^{iqx},\nn
\hD_0(q)&=&\int dx\hD_0(x,0)e^{iqx},
\eea
given by
\be\label{frmprop}
\hG_{0\tau}(q)=(q\br+m_\tau)\left[\begin{pmatrix}\frac1{q^2-m_\tau^2+i\epsilon}&2\pi i\delta(q^2-m^2_\tau)\Theta(-q^0)\cr2\pi i\delta(q^2-m^2_\tau)\Theta(q^0)&-\frac1{q^2-m^2_\tau-i\epsilon}\end{pmatrix}+2\pi i\delta(q^2-m^2)n_{q\tau}\begin{pmatrix}1&-1\cr-1&1\end{pmatrix}\right]
\ee
where the occupation number at zero temperature and finite density is given by
\be\label{occnum}
n_{q\tau}=\Theta(q^0)\Theta(\mu_\tau-\epsilon_{\v{q}\tau})+\Theta(-q^0)\Theta(-\epsilon_{\v{q}\tau}-\mu_\tau)
\ee
with $\mu_e=\mu$, $\mu_p=0$ and $\epsilon_{\v{q}\tau}=\sqrt{m_\tau^2+\v{q}^2}$ \cite{maxwell}. The photon propagator is used in Feynman gauge, $\hD_0^{\mu\nu}(q)=g^{\mu\nu}\hD_0(q)$ with
\be\label{scprop}
\hD_0(q)=\begin{pmatrix}\frac1{q^2+i\epsilon}&-2\pi i\delta(q^2)\Theta(-q^0)\cr-2\pi i\delta(q^2)\Theta(q^0)&-\frac1{q^2-i\epsilon}\end{pmatrix}
\ee

The structure of these matrices reflects a general rule, valid for any local composite operators,
\be\label{spropctp}
i\hD(x,y)=\begin{pmatrix}\la T[\phi(x)\phi(y)]\ra&\la\phi(y)\phi(x)\ra\cr\la\phi(x)\phi(y)\ra&\la T[\phi(y)\phi(x)]\ra^*\end{pmatrix}
=i\begin{pmatrix}D^n+iD^i&-D^f+iD^i\cr D^f+iD^i&-D^n+iD^i\end{pmatrix}
\ee
for bosonic operators and 
\be
i\hG(x,y)=\begin{pmatrix}\la0|T[\psi(x)\psib(y)]|0\ra&\la0|\psib(y)\psi(x)]|0\ra\cr-\la0|\psi(x)\psib(y)|0\ra&\la0|T[(\gamma^0\psi(y))((\psib(x)\gamma^0)]|0\ra^*\end{pmatrix}
=i(\partial\br_x+m)\begin{pmatrix}G^n+iG^i&G^f-iG^i\cr-G^f-iG^i&-G^n+iG^i\end{pmatrix}
\ee
for fermionic operators. Furthermore, $D^n(q)=D^n(-q)$, $D^i(q)=D^i(-q)$, $G^n(q)=G^n(-q)$, $G^i(q)=G^i(-q)$ are real and  $D^f(q)=-D^f(-q)$ and $G^f(q)=-G^f(-q)$ are imaginary. Note that the off-shell contributions are in the real part of the diagonal blocks only. The environment contributions are always on shell. In case the photon propagator $D^n$ and $D^f$ are the near and far field, 
\be\label{raprop}
D^{\stackrel{r}{a}}=D^n\pm D^f
\ee
give the retarded and advanced Green functions of classical electrodynamics.

We return now to the generator functional \eq{wpint} which assumes the form 
\be\label{genfqed}
e^{iW[\hj,\hetab_e,\heta_e,\hetab_p,\heta_p]}=\exp\left[i\sum_{\sigma=\pm1}\sigma\int dx\fd{}{j^{\sigma\mu}(x)}\sum_\tau e_\tau\fd{}{\eta^\sigma_\tau(x)}\gamma^\mu\fd{}{\etab^\sigma_\tau(x)}\right]
e^{iW_0[\hj,\hetab_e,\heta_e,\hetab_p,\heta_p]}.
\ee
The lowest order contribution to the transition probability \eq{trprobfd} comes from the $\ord{\etab_p^4\eta_p^4\etab_e^4\eta_e^4e^4}$ term
\bea\label{ethord}
&&\frac{(ie)^4}{4!}\left[\sum_{\sigma=\pm1}\sigma\int dx\fd{}{j^{\sigma\mu}(x)}\sum_\tau e_\tau\fd{}{\eta^\sigma_\tau(x)}\gamma^\mu\fd{}{\etab^\sigma_\tau(x)}\right]^4
\frac{(-i)^2}{2^22!}\left[\int dxdy\hat j^\mu(x)\hD_{0\mu\nu}(x,y)\hat j^\nu(y)\right]^2\nn
&&\times\frac{(-i)^4}{4!}\left[\int dxdy\hetab_e(x)\hG_{0e}(x,y)\heta_e(y)\right]^4\frac{(-i)^4}{4!}\left[\int dxdy\hetab_p(x)\hG_{0p}(x,y)\heta_p(y)\right]^4
\eea
in Eq. \eq{genfqed} where we can use $Z_e=Z_p=1$. 

Our system consists of the colliding electron-proton pair and photons, their environment being the remaining electrons in the gas with occupation number \eq{occnum}. The operators of the free equation of motion that truncate the Green function in Eq. \eq{trprobfd} when they act on the external leg according to the reduction formulas extract the residuum at the mass shell. Owing to the identity 
\be\label{distrz}
\delta(x)x=0
\ee
this operation suppresses all mass-shell contribution, such as the off-diagonal CTP blocks and the $\ord{n_\v{k}}$ environment piece of the electron propagator \eq{frmprop}. In physical terms, we find the usual OTP propagator handling our system, whose asymptotic sector contains the colliding particles only. Naturally, the soft photon component of asymptotic charged states \cite{bloch}, ignored in this work, should be constructed as in the usual scattering processes without environment.

\begin{figure}
\parbox{8cm}{\includegraphics[scale=.4]{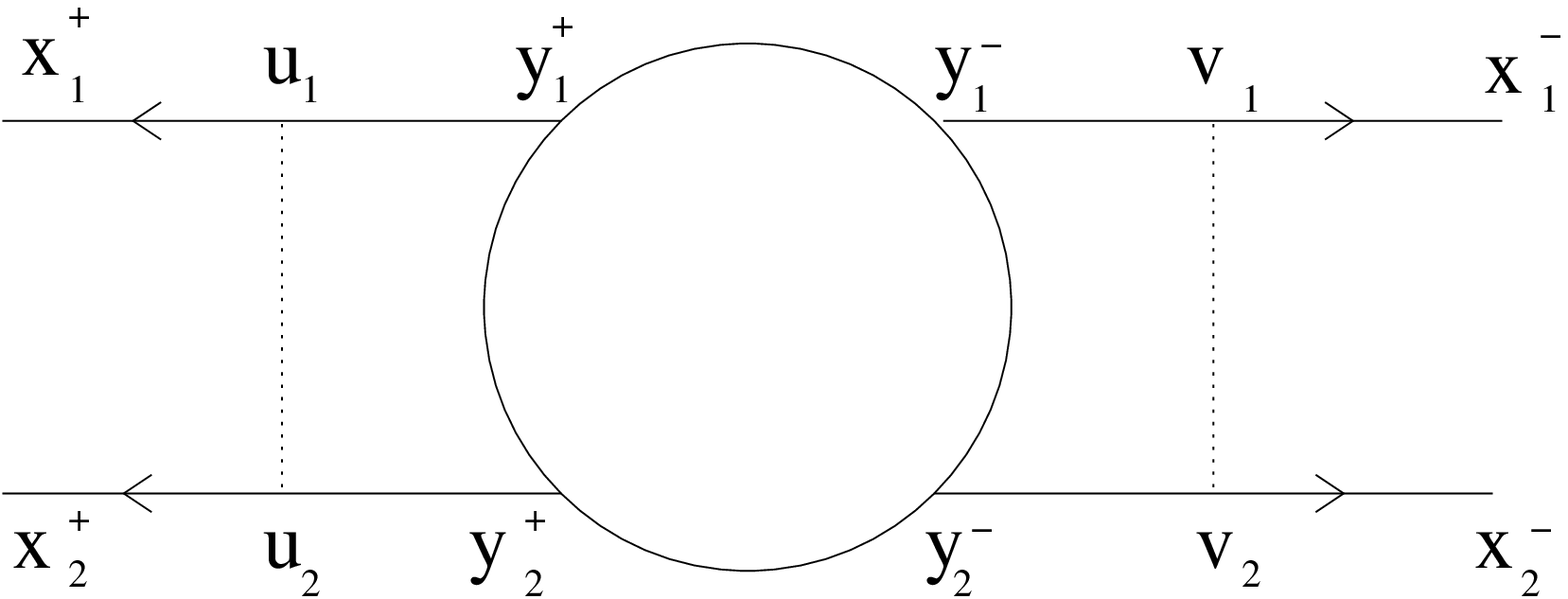}\\(a)}
\parbox{8cm}{\includegraphics[scale=.4]{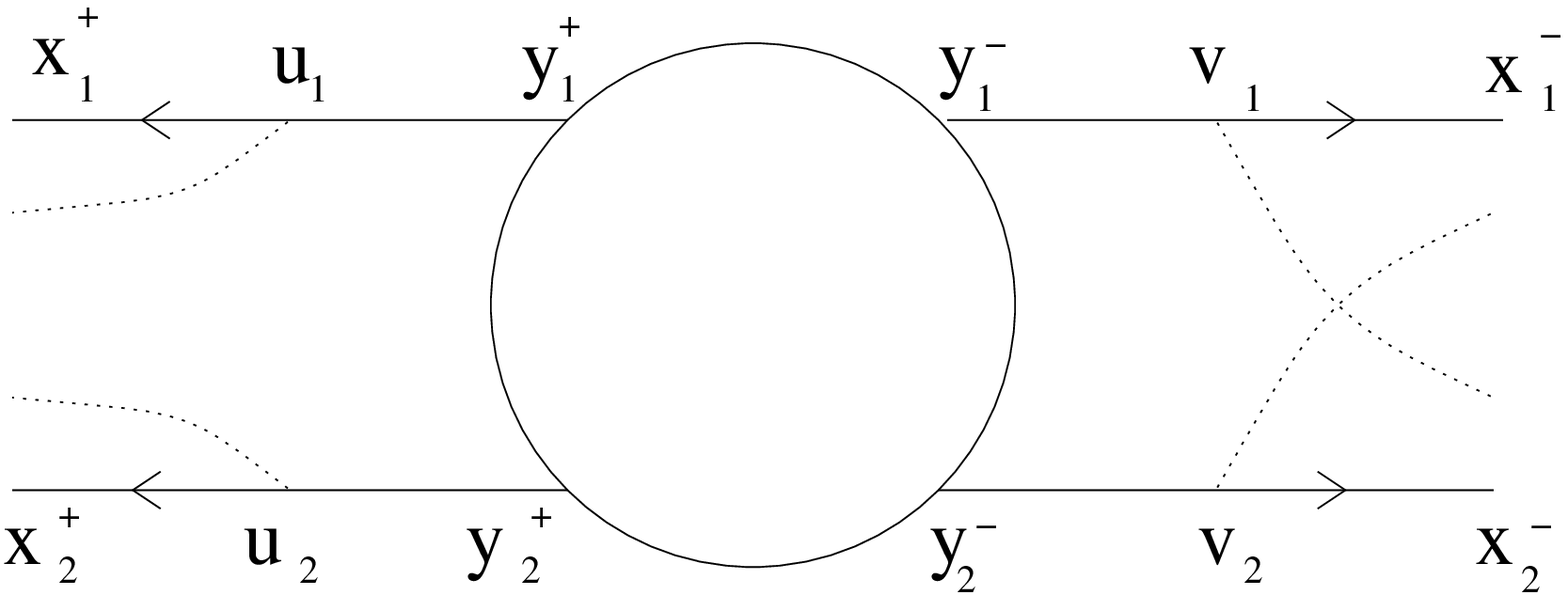}\\(b)}
\caption{The leading order graphs contributing to the transition probability: (a): direct, (b): entanglement contributions. The circle represents the initial density matrix. The CTP-diagonal block of the photon propagator connects charged particle lines on the same time axis in graph (a); the off-diagonal blocks couple the two time axes and their photon lines traverse the final time, the left and right end of graph (b), with periodic boundary conditions in time, i.e., lines ending at the same altitude at the left and right end of the graph are joined.}\label{ctpgraphs}
\end{figure}

There are two topologically different graphs characterizing the contributions in \eq{ethord}, they are depicted in Fig. \ref{ctpgraphs}. Graph (a),
\bea\label{direct}
&&-e^4\int du_1du_2D^{++}_{0\mu\nu}(u_1-u_2)G^{++}_{c^+c'^+}(y^+_1-u_1)\gamma^\mu_{c'^+a'^+}G^{++}_{a'^+a^+}(u_1-x^+_1)G^{++}_{d^+d'^+}(y^+_2-u_2)\gamma^\nu_{d'^+b'^+}G^{++}_{b'^+b^+}(u_2-x^+_2)\nn
&&\times\int dv_1dv_2D^{--}_{0\mu'\nu'}(v_1-v_2)G^{--}_{c^-c'^-}(y^-_1-v_1)\gamma^{\mu'*}_{c'^-a'^-}G^{--}_{a'^-a^-}(v_1-x^-_1)G^{--}_{d^-d'^-}(y^-_2-v_2)\gamma^{\nu'*}_{d'^-b'^-}G^{--}_{b'^-b^-}(v_2-x^-_2).
\eea
contains the CTP-diagonal contributions of the photon propagator and displays interaction between the colliding charges by exchanging a photon. The state of the colliding particles remains pure after interacting and the transition probability is the amplitude times its complex conjugate. Graph (b) in Fig. \ref{ctpgraphs},
\bea\label{entprob}
&&-e^4\int du_1du_2D^{+-}_{0\mu\nu'}(u_1-v_2)G^{++}_{c^+c'^+}(y^+_1-u_1)\gamma^\mu_{c'^+a'^+}G^{++}_{a'^+a^+}(u_1-x^+_1)G^{++}_{d^+d'^+}(y^+_2-u_2)\gamma^\nu_{d'^+b'^+}G^{++}_{b'^+b^+}(u_2-x^+_2)\nn
&&\times\int dv_1dv_2D^{-+}_{0\mu'\nu}(v_1-u_2)G^{--}_{c^-c'^-}(y^-_1-v_1)\gamma^{\mu'*}_{c'^-a'^-}G^{--}_{a'^-a^-}(v_1-x^-_1)G^{--}_{d^-d'^-}(y^-_2-v_2)\gamma^{\nu'*}_{d'^-b'^-}G^{--}_{b'^-b^-}(v_2-x^-_2),
\eea
includes the CTP off-diagonal part of the photon propagator. Note that these blocks contain the one-shell Wightman function and have to be represented by a line of the corresponding particle, namely, the photon that connects the two time axes by passing between them at the final time. The graphs, therefore, are equipped with periodic boundary conditions in time; lines reaching the final time at one time axis enter at the other time axis on the other side of the diagram. Therefore, this graph describes the contributions of final states containing two photons to the scattering. When a photon is attached to the same particle at a different time axis, then this contribution represents the soft photon component of the colliding charges. But, we ignore forward scattering and the leading-order graph must contain photons that are exchanged between the colliding particles. Since the support of the off-diagonal CTP blocks of the free photon propagator is on the mass shell, the emission of a single photon is kinematically excluded and this unusual graph is vanishing in the absence of the electron gas.

\subsection{t-channel resummation}\label{tchann}
The higher-order corrections to the transition probability \eq{trprobfd} can be partially resummed by means of the skeleton expansion, where graphs with different self-energy or vertex insertions are resummed. The result is a graph similar to the lowest-order contribution except that the propagators and vertices are replaced by their exact expressions. Such a resummation represents an important improvement for kinematical regimes where one of these quantities assumes values strongly different than in leading order. We shall be interested in scattering process where collective particle-hole excitations take place in the t channel. For this end, the usual resummation of the photon self-energy is necessary.

A natural consistency requirement for any partial resummation is to preserve gauge invariance to properly separate the physical sector from the IR singular, nonphysical gauge 
modes. A simple way of checking gauge invariance of a given calculation of the irreducible vertex functions is to construct the effective action and verify its invariance under gauge transformation. Photons are electrically neutral therefore their effective CTP action obtained by the partial resummation of the self-energy $\hat\Sigma$ is the free Maxwell action
\be
\hf\int dxdy\hat A_\mu(x)\hD_0^{-1\mu\nu}(x,y)\hat A_\nu(y)
\ee
except that the inverse free photon propagator $\hD^{-1}_0$ is replaced by the dressed one,
\be\label{resumprop}
\hD^{-1}=\hD^{-1}_0-\hat\Sigma.
\ee
The self-energy is transverse, $\partial^\mu\hat\Sigma_{\mu\nu}(x,y)=0$ and our partial resummation is consistent. 

\begin{figure}
\parbox{8cm}{\includegraphics[scale=.4]{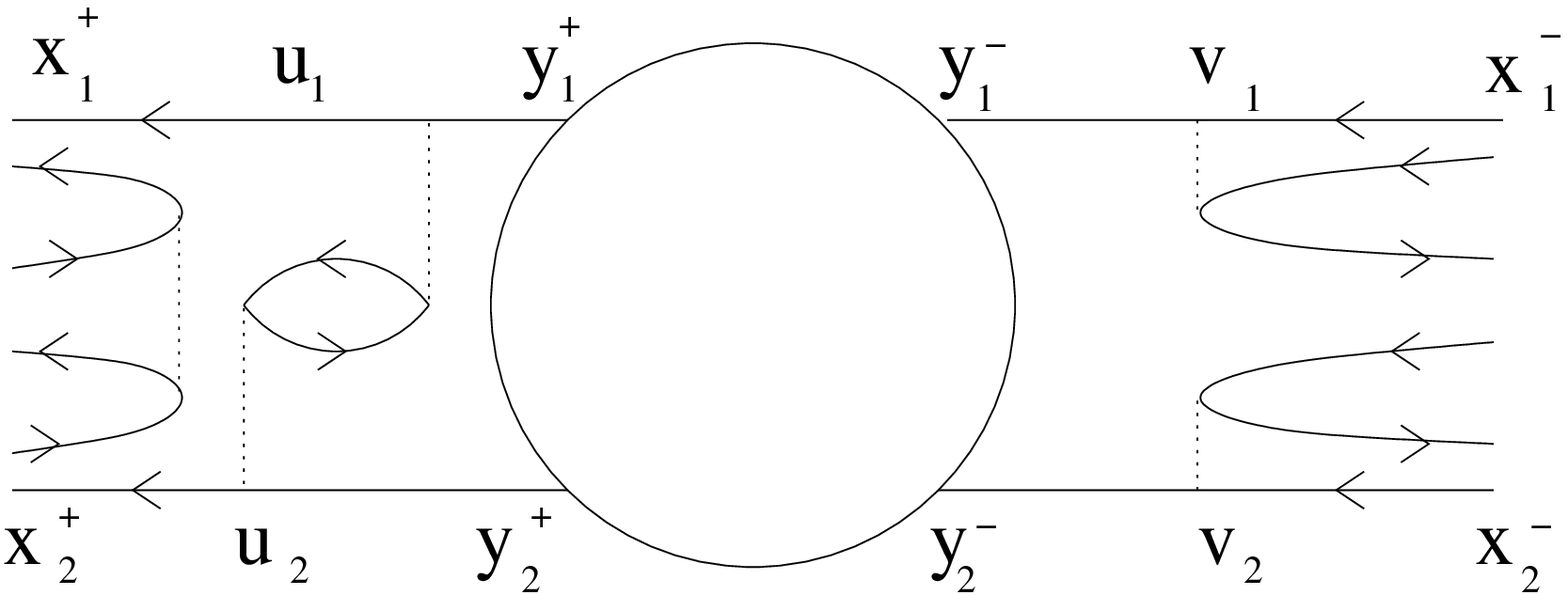}\\(a)}
\parbox{8cm}{\includegraphics[scale=.4]{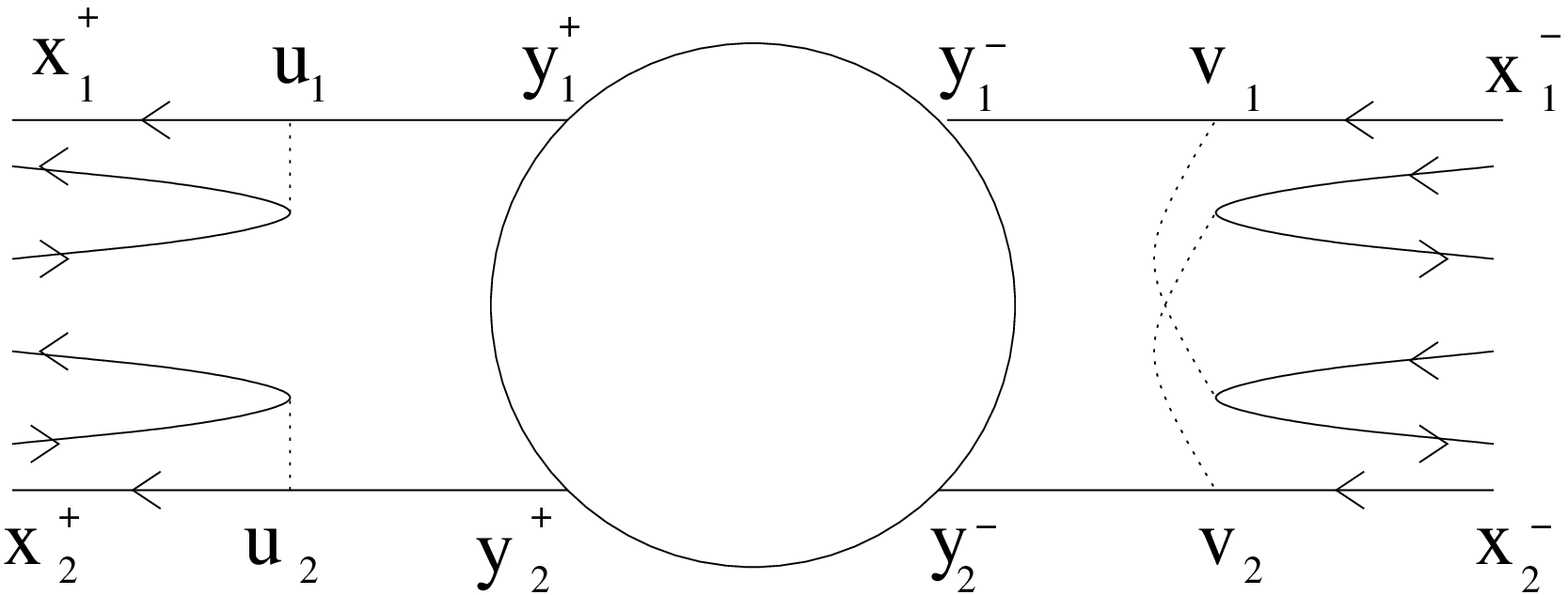}\\(b)}
\caption{Two higher order graphs contributing to the transition probability: (a): direct, (b): entanglement contributions. The particle-hole pairs dress the propagation of the photon in each time axis separately in the direct graph and traverse the final time in the entangled graph.}\label{ctpgraphspi}
\end{figure}

The reduction formula for the transition probability \eq{trprobfd} yields the sum
\be\label{trprobs}
P(f\leftarrow i)=P_d+P_e,
\ee
where the first and the second term corresponds to expressions \eq{direct} and \eq{entprob}, respectively. The resummation of the phonon-self-energy amounts to the replacement of the free photon propagator with \eq{resumprop} in Eqs. \eq{direct} and \eq{entprob}. The CTP-diagonal direct term,
\bea\label{pdir}
P_d&=&e^4\prod_{\sigma=\pm}\int dx^\sigma_1dx^\sigma_2dy^\sigma_1dy^\sigma_2e^{iq_1(y^+_1-y^-_1)+iq_2(y^+_2-y^-_2)-ip_1(x^+_1-x^-_1)-ip_2(x^+_2-x^-_2)}\nn
&&\times\ub_{1f}(i\partial\br_{y^+_1}-m)\ub_{2f}(i\partial\br_{y^+_2}-M)\ub^*_{1f}(-i\partial\br_{y^-_1}-m)\ub^*_{2f}(-i\partial\br_{y^-_2}-M)\nn
&&\times\int du_1du_2D^{++}_{\mu\nu}(u_1-u_2)G^{++}_{c^+c'^+}(y^+_1-u_1)\gamma^\mu_{c'^+a'^+}G^{++}_{a'^+a^+}(u_1-x^+_1)G^{++}_{d^+d'^+}(y^+_2-u_2)\gamma^\nu_{d'^+b'^+}G^{++}_{b'^+b^+}(u_2-x^+_2)\nn
&&\times\int dv_1dv_2D^{--}_{\mu'\nu'}(v_1-v_2)G^{--}_{c^-c'^-}(y^-_1-v_1)\gamma^{\mu'*}_{c'^-a'^-}G^{--}_{a'^-a^-}(v_1-x^-_1)G^{--}_{d^-d'^-}(y^-_2-v_2)\gamma^{\nu'*}_{d'^-b'^-}G^{--}_{b'^-b^-}(v_2-x^-_2)\nn
&&\times(-i\partial\br_{x^+_1}-m)u_{1i}(-i\partial\br_{x^+_2}-M)u_{2i}(i\partial\br_{x^-_1}-m)u^*_{1i}(i\partial\br_{x^-_2}-M)u^*_{2i},
\eea
contains the contributions of the colliding particle interaction. Since the support of the off-diagonal CTP blocks $D_0^{\pm\mp}$ is the mass-shell all free photon lines in the Schwinger-Dyson resummed graphs stand for $D_0^{\pm\pm}$. Hence, the number of $\Sigma^{\pm\mp}$ insertions is always even in $P_d$. An $\ord{\Sigma^{++}(\Sigma^{+-})^2}$ piece of the geometrical series, resummed Eq. \eq{resumprop}, is shown in Fig. \ref{ctpgraphspi} (a). The self-energy insertion $\Sigma^{++}$ describes a particle-hole pair creation contributing to the interactions of the colliding particles on the $+$ axis. The correlation of the two colliding particles on the $-$ axis represents interaction. In fact, each particle interacts with a particle-hole pair that does not interact with each other but forms a final state that is matched with two particle-hole pairs of the other time axis. These pairs are the source of correlation because they interact by arising from the same photon. Therefore, the matching of the asymptotic states at the final time can transfer correlations from one time axis to the other one. Note that the off-diagonal CTP blocks of the photon self-energy, $\Sigma^{\pm\mp}$, have a special role in the transition probability; they represent the asymptotic particle-hole state contributions, the backreaction of the colliding charge on the electron gas. For instance, the $n$-fold insertion of the one-loop $\Sigma^{\pm\mp}$ captures the dynamics of $n$ asymptotic particle-hole pairs. 

The contributions of the CTP off-diagonal blocks of the photon propagator,
\bea\label{pent}
P_e&=&e^4\prod_{\sigma=\pm}\int dx^\sigma_1dx^\sigma_2dy^\sigma_1dy^\sigma_2e^{iq_1(y^+_1-y^-_1)+iq_2(y^+_2-y^-_2)-ip_1(x^+_1-x^-_1)-ip_2(x^+_2-x^-_2)}\nn
&&\times\ub_{1f}(i\partial\br_{y^+_1}-m)\ub_{2f}(i\partial\br_{y^+_2}-M)\ub^*_{1f}(-i\partial\br_{y^-_1}-m)\ub^*_{2f}(-i\partial\br_{y^-_2}-M)\nn
&&\times\int du_1du_2D^{+-}_{\mu\nu'}(u_1-v_2)G^{++}_{c^+c'^+}(y^+_1-u_1)\gamma^\mu_{c'^+a'^+}G^{++}_{a'^+a^+}(u_1-x^+_1)G^{++}_{d^+d'^+}(y^+_2-u_2)\gamma^\nu_{d'^+b'^+}G^{++}_{b'^+b^+}(u_2-x^+_2)\nn
&&\times\int dv_1dv_2D^{-+}_{\mu'\nu}(v_1-u_2)G^{--}_{c^-c'^-}(y^-_1-v_1)\gamma^{\mu'*}_{c'^-a'^-}G^{--}_{a'^-a^-}(v_1-x^-_1)G^{--}_{d^-d'^-}(y^-_2-v_2)\gamma^{\nu'*}_{d'^-b'^-}G^{--}_{b'^-b^-}(v_2-x^-_2)\nn
&&\times(-i\partial\br_{x^+_1}-m)u_{1i}(-i\partial\br_{x^+_2}-M)u_{2i}(i\partial\br_{x^-_1}m)u^*_{1i}(i\partial\br_{x^-_2}-M)u^*_{2i},
\eea
reflect colliding charge-gas entanglement. An $\ord{(\Sigma^{+-})^2}$ graph contributing to $P_e$ is depicted in Fig. \ref{ctpgraphspi} (b). The colliding charges produce particle-hole pairs on both time axes. The matching of these pairs at the final time establishes a coupling between the time axes which qualifies entanglement according to the remarks made after Eq. \eq{inflfunc} .

A straightforward calculation gives the transition probabilities 
\bea\label{treprobde}
P_d&=&V^{(4)}(2\pi)^4\delta(q_1-p_1+q_2-p_2)|{\cal T}_d|^2,\nn
P_e&=&V^{(4)}(2\pi)^4\delta(q_1-p_1+q_2-p_2)|{\cal T}_e|^2,
\eea
where $V^{(4)}$ is the four-volume and the ``transition amplitude'' squares are given by positive semidefinite expressions
\bea\label{amplsqde}
|{\cal T}_d|^2&=&e^4|(\ub_{1f}\gamma^\mu u_{1i})(\ub_{2f}\gamma^\nu u_{2i})D^{++}_{\mu\nu}(q_1-p_1)|^2,\nn
|{\cal T}_e|^2&=&e^4|(\ub_{1f}\gamma^\mu u_{1i})(\ub_{2f}\gamma^\nu u_{2i})^*D^{+-}_{\mu\nu}(q_1-p_1)|^2,
\eea
by means of the relations $D_{\mu\nu}^{\sigma,\sigma'}(p)=[D_{\nu\mu}^{-\sigma,-\sigma'}(-p)]^*$ for the CTP propagator for any, not necessarily free, local four-vector operator.

\subsection{Photon self-energy}
Before presenting the photon self-energy let us work out two useful parameterizations of Lorentz tensors appearing in this work. The photon self-energy $\Sigma^{\mu\nu}$ and propagator $D^{\mu\nu}$ have two intrinsic Lorentz vectors, the four-momentum $q^\mu$ and a unit timelike vector, assumed to be $u^\mu=(1,\v{0})$ in the rest frame of the gas. The symmetric tensor $M$, denoting either self-energy or propagator, constructed by the help of these vectors has three free scalar parameters, for instance the coefficients of the terms $q^\mu q^\nu$, $n^\mu n^\nu$, and  $q^\mu n^\nu+n^\mu q^\nu$. Current conservation allows us to ignore the longitudinal part of the photon propagator in Eqs. \eq{amplsqde} hence transversality, $q_\mu M^{\mu\nu}(q)=0$, reduces the number of free scalars to two that will be chosen $M_g=M^{\mu\nu}g_{\nu\mu}$ and $M_u=u_\mu M^{\mu\nu}u_\nu$. This parameterization is useful to calculate loop integrals, cf. Eq. \eq{olselfene} below.

When the inverse of a tensor is needed for the calculation of the propagator by means of the Schwinger-Dyson equation \eq{resumprop}, then the parameters $M_g$ and $M_u$ are not useful anymore. Instead, it is more advantageous to split the four-dimensional transverse subspace, identified by the projector $T^{\mu\nu}=g^{\mu\nu}-q^\mu q^\nu/q^2$ into three-dimensional transverse and longitudinal subspaces, $T=P_\ell+P_t$, where the projectors 
\bea
P^{\mu\nu}_t&=&-\begin{pmatrix}0&0\cr0&\v{T}\end{pmatrix},\nn
P^{\mu\nu}_\ell&=&\frac1{1-\nu^2}\begin{pmatrix}1&\v{n}\nu\cr\v{n}\nu&\nu^2\v{L}\end{pmatrix}
\eea
corresponding to the transverse and longitudinal modes, respectively, in Lorentz gauge are given in terms of the three-dimensional longitudinal and transverse projectors, $\v{L}=\v{n}\otimes\v{n}$, $\v{T}=\openone-\v{L}$, with $\v{n}=\v{q}/|\v{q}|$ and $
\nu=q^0/|\v{q}|$. The inverse of the matrix 
\be\label{lbpar}
M^{\mu\nu}(M_\ell,M_t)=M_\ell P^{\mu\nu}_\ell+M_tP^{\mu\nu}_t
\ee
can be obtained in an obvious manner within both subspaces, $M^{-1}(M_\ell,M_t)=M^{\mu\nu}(1/M_\ell,1/M_t)$. It is easy to check the relation
\be\label{tworules}
M_\ell=(1-\nu^2)M_u,~~~M_t=\hf[M_g+(\nu^2-1)M_u]
\ee
between the two different parametrization.

The one-loop photon self-energy in an environment \cite{kislinger} has already been studied thoroughly; we need its CTP form \cite{weldonse} for a zero temperature nondegenerate electron gas of Fermi momentum $\kfer$,
\be\label{olselfene}
\Sigma^{(\sigma\mu)(\sigma'\nu)}(q)=\Sigma^{\sigma\sigma'}_u(q)\begin{pmatrix}1&\v{n}\nu\cr\v{n}\nu&\nu^2\v{L}\end{pmatrix}+\hf[\Sigma^{\sigma\sigma'}_u(q)(1-\nu^2)-\Sigma_g^{\sigma\sigma'}(q)]\begin{pmatrix}0&0\cr0&\v{T}\end{pmatrix}.
\ee
The $2\times2$ CTP blocks are given by the matrices which can be written as sums of the vacuum and electron gas contributions as \cite{maxwell}
\be
\hat\Sigma_x=\hat\Sigma_{x~vac}+\hat\Sigma_{x~gas}
\ee
where $x$ stands for the letter $g$ or $u$ and the self-energy displays the structure
\be\label{selfenctp}
\hat\Sigma=\hat\sigma\begin{pmatrix}\Pi^n+i\Pi^i&-\Pi^f+i\Pi^i\cr\Pi^f+i\Pi^i&-\Pi^n+i\Pi^i\end{pmatrix}\hat\sigma
\ee
with
\be
\hat\sigma=\begin{pmatrix}1&0\cr0&-1\end{pmatrix}.
\ee
The vacuum contributions contain the functions
\bea
\Pi^n_{vac}(q)&=&\frac{\alpha}{3\pi}q^2\left\{\frac13+2\left(1+\frac{2m^2}{q^2}\right)\left[\sqrt{\frac{4m^2}{q^2}-1}~\mr{arccot}\sqrt{\frac{4m^2}{q^2}-1}-1\right]\right\}\nn
\Pi^f_{vac}(q)&=&-i\alpha\left(m^2-\frac{q^2}2\right)\frac{4\Theta(-q^0-m)}{3|\v{q}|}\int_0^{\sqrt{q^{02}-m^2}}dp\frac{p}{\omega(p)}\Theta(2p|\v{q}|-|q^2+2\omega(p)q^0|)
\eea
with $\alpha=e^2/4\pi$ and $\omega(q)=\sqrt{m^2+\v{p}^2}$,
\bea\label{xvac}
\Pi^n_{g~vac}(q)&=&3\Re\Pi^n_{vac}(q),\nn
\Pi^n_{u~ vac}(q)&=&\frac{\Re\Pi^n_{vac}(q)}{1-\nu^2},
\eea
and similar expression for $\Pi^f_{x~vac}$. We shall consider below the nonrelativistic regime, $|q^0|<m$ where $\Pi^f_{vac}(q)=0$ and $\Im\Pi^n_{vac}(q)=0$. 

A zero temperature nonrelativistic gas of electrons with Fermi momentum $\kfer\ll m$ gives
\bea\label{reimx}
\Pi^n_{x~gas}(q)&=&\tilde\Pi_{x~gas}^n(q)+\tilde\Pi_{x~gas}^n(-q),\nn
\Pi^f_{x~gas}(q)&=&-i[\tilde\Pi_{x~gas}^f(q)-\tilde\Pi_{x~gas}^f(-q)],\nn
\Pi^i_{x~gas}(q)&=&\tilde\Pi_{x~gas}^i(q)+\tilde\Pi_{x~gas}^i(-q),
\eea
with
\bea\label{abselfen}
\tilde\Pi_{g~gas}^n(q)&=&\frac{2\alpha \kfer^2m}{\pi|\v{q}|}\left(1+\frac{q^2}{2m^2}\right)L(q),\nn
\tilde\Pi_{u~gas}^n(q)&=&\frac{2\alpha \kfer^2m}{\pi|\v{q}|}\left(1+\frac{q^2}{4m^2}+\frac{q^0}{m}\right)L(q),\nn
\tilde\Pi_{g~gas}^f(q)&=&-\frac{\alpha \kfer^2m}{|\v{q}|}\left(1+\frac{q^2}{2m^2}\right)M(q),\nn
\tilde\Pi_{u~gas}^f(q)&=&-\frac{\alpha \kfer^2m}{|\v{q}|}\left(1+\frac{q^2}{4m^2}+\frac{q^0}{m}\right)M(q),\nn
\tilde\Pi_{g~gas}^i(q)&=&-\frac{\alpha \kfer^2m}{|\v{q}|}\left(1+\frac{q^2}{2m^2}\right)N(q),\nn
\tilde\Pi_{u~gas}^i(q)&=&-\frac{\alpha \kfer^2m}{|\v{q}|}\left(1+\frac{q^2}{4m^2}+\frac{q^0}{m}\right)N(q),
\eea
and 
\bea\label{lmselfen}
L(q)&=&r+\hf(1-r^2)\ln\left|\frac{r+1}{r-1}\right|,\nn
M(q)&=&\Theta(1-|r|)(1-r^2),\nn
N(q)&=&\begin{cases}1-r^2&|\v{q}|>2\kfer,~-1<r<1\cr1-r^2&|\v{q}|<2\kfer,~-1-\frac{|\v{q}|}{\kfer}<r<1\cr
\frac{(q^0+2m)q^0}{\kfer^2}&|\v{q}|<2\kfer,~-\frac{|\v{q}|}{2\kfer}<r<1-\frac{|\v{q}|}{\kfer}\end{cases}.
\eea
The $q$ dependence of the functions $L(q)$, $M(q)$ and $N(q)$ is given through the combination $r=(q^2+2mq^0)/2|\v{q}|\kfer$.

\section{Photon propagator}\label{photprop}
The actual resummation of the geometrical series with the self-energy is carried out by the algebraic solution of the Schwinger-Dyson-Kadanoff-Baym equation, Eq. \eq{resumprop}. We first consider the CTP and after that the Lorentz tensor structure of the propagator.

\subsection{CTP blocks}
The CTP blocks structure of the propagator is shown in Eq. \eq{spropctp} and its inverse, both $\hD_0^{-1}$ and $\hat\Pi$, have a common, slightly different $2\times2$ block structure of Eq. \eq{selfenctp}, related by the change of basis by $\hat\sigma$. The Green-function-like structure is inherited during multiplications where $\hat\sigma$ is treated a metric tensor, i.e., the Green function and its inverse are interpreted as covariant and contravariant tensors, respectively. In fact, one can easily prove the relations
\bea\label{lemme}
(D_1\hat\sigma D_2\hat\sigma\cdots D_j)^{\stackrel{r}{a}}
&=&D^{\stackrel{r}{a}}_1D^{\stackrel{r}{a}}_2\cdots D^{\stackrel{r}{a}}_j,\nn
(D_1\hat\sigma D_2\hat\sigma\cdots D_j)^i
&=&(D_1\hat\sigma D_2\hat\sigma\cdots D_{j-1})^iD_j^a+(D_1\hat\sigma D_2\hat\sigma\cdots D_{j-1})^rD_j^i,
\eea
for a set of Green function like $2\times2$ matrices, $\hD_1,\ldots,\hD_n$ by recursion \cite{maxwell}. These relations are useful when the propagator \eq{resumprop} is written as
\be\label{resumprops}
\hD=\frac1{1-\hD_0\hat\sigma\hat\Pi\hat\sigma}\hat D_0
\ee
and the first equation in \eq{lemme} gives
\be\label{resumra}
D^{\stackrel{r}{a}}=\frac1{1-D_0^{\stackrel{r}{a}}\Pi^{\stackrel{r}{a}}}D_0^{\stackrel{r}{a}}
=\frac1{D_0^{-1}-\Pi^{\stackrel{r}{a}}}.
\ee
The imaginary part is obtained by using the second equation in \eq{lemme},
\bea
\left(\frac1{1-\hD_0\hat\sigma\hat\Pi\hat\sigma}\right)^i
&=&\sum_{j=1}^\infty[(D_0\sigma\Pi\sigma)^j]^i(D_0\sigma\Pi\sigma)^a
+\sum_{j=0}^\infty[(D_0\sigma\Pi\sigma)^j]^r(D_0\sigma\Pi\sigma)^i\nn
&=&\left(\frac1{1-\hD_0\hat\sigma\hat\Pi\hat\sigma}\right)^iD_0^a\Pi^a+\frac1{1-D_0^r\Pi^r}(D_0\sigma\Pi\sigma)^i,
\eea
yielding
\be
\left(\frac1{1-\hD_0\hat\sigma\hat\Pi\hat\sigma}\right)^i
=\frac1{1-D_0^r\Pi^r}(D_0^i\Pi^a+D_0^r\Pi^i)\frac1{1-D_0^a\Pi^a}
\ee
and
\be\label{resumi}
D^i=D^r\Pi^iD^a
\ee
where the equation $D_0^iD_0^{-1}=0$, an application of Eq. \eq{distrz}, has been used. Finally, Eqs. \eq{resumra} and \eq{resumi} together with \eq{spropctp} and \eq{raprop} give the CTP structure of the resummed photon propagator.

\subsection{Lorentz blocks}
Next, we work out the retarded and advanced CTP blocks of the Lorentz tensors appearing on the right-hand side of the Schwinger-Dyson equation, \eq{resumprop}. The inverse free propagator is usually obtained either by the identification of the kernel of the quadratic part of the action or simply by the inversion of the propagator in momentum space. Neither of these ways is trivial in the CTP scheme. The two time axes are coupled at the final time in the path integral \eq{wpint}. One cannot even have an inverse propagator that is invariant under translation in time. The CTP propagators can nevertheless be derived in a straightforward but rather lengthy way in the path integral formalism by carefully implementing the boundary conditions in time for free fields at finite $\Delta t$ and performing the limit $\Delta t\to0$, followed by $t_f-t_i\to\infty$. A translation-invariant kernel and diagonal form in frequency space is recovered without problem after the second limiting procedure. 

A shorter and more promising way is to derive the propagators \eq{frmprop} and \eq{scprop} in the operator formalism where they are automatically diagonal in the Fourier space in the limit $t_f-t_i\to\infty$. But, the difficulties are encountered in this case when the inverse of distributions, handling the mass-shell singularities, is sought. A simple, natural way out is to regulate the distributions by the replacement 
\be
\delta(q^2)=\frac\epsilon\pi\frac1{q^2+\epsilon^2}
\ee
and performing the inversion with small but finite $\epsilon$. The result one finds in this manner is
\be\label{invctpprop}
\hD_0^{-1}(q)=q^2\begin{pmatrix}1&0\cr0&-1\end{pmatrix}+i\epsilon\begin{pmatrix}1&-2\Theta(-q^0)\cr-2\Theta(q^0)&1\end{pmatrix}.
\ee
Note the transmutation of the coupling between the two time axes. One the one hand, this coupling is $\epsilon$-independent, $\ord{\epsilon^0}$ and is localized in time at $t=t_f$ in the action written in the space-time for finite $t_f-t_f$, like in Eq. \eq{ctpact}. On the other hand, when we rewrite the action in the momentum space after the limit $t_f-t_i\to\infty$ by means of the free inverse propagator \eq{invctpprop} then the boundary conditions in time are difficult to trace and the coupling between the time axes is represented by a weak, $\ord{\epsilon}$ term in the free action which acts during the whole time evolution in a time-independent manner.

The retarded and advanced part of the self-energy \eq{olselfene} are
\be
\Pi^{\stackrel{r}{a}}_\ell=(1-\nu^2)\Pi_u^{\stackrel{r}{a}},~~~
\Pi^{\stackrel{r}{a}}_t=\hf[\Pi_g^{\stackrel{r}{a}}+(\nu^2-1)\Pi_u^{\stackrel{r}{a}}]
\ee
according to \eq{tworules} where $\Pi^{\stackrel{r}{a}}=\Pi^n\pm\Pi^f$. An important feature of the self-energy component $\Pi^i$ is that it is always nonpositive and we can in this manner ignore the second, $\ord{\epsilon}$ term of the free propagator \eq{invctpprop} beside the self-energy, leading to $D^{-1n}_0=-q^2(P_\ell+P_t)$ and $D^{-1f}_0=D^{-1i}_0=0$.

The Schwinger-Dyson resummed retarded and advanced propagators are found first by means of Eq. \eq{resumra},
\be\label{retadvpr}
D_\ell^{\stackrel{r}{a}}=\frac1{(\nu^2-1)(\v{q}^2-\Pi_u^{\stackrel{r}{a}})},~~~
D_t=\frac1{\hf\Pi_g^{\stackrel{r}{a}}+(\nu^2-1)(\hf\Pi_u^{\stackrel{r}{a}}+\v{k}^2)}.
\ee
The imaginary part of the propagator, given by Eq. \eq{resumi} turns out to be
\be
D^i_\ell=D_\ell^rD_\ell^a\Pi_u^i(1-\nu^2),~~~D^i_t=\hf D_t^rD_t^a[\Pi_g^i+(\nu^2-1)\Pi_u^i].
\ee
This propagator inserted in Eqs. \eq{treprobde}-\eq{amplsqde} gives the partially resummed expression for the transition probability.

\subsection{Nonrelativistic spectral representation}
We close this section by a brief summary of the spectral representation without relativistic symmetries for the two-point functions of local transverse vector operators, in particular for the photon field or the electric current within the CTP scheme. The spectral representation of a relativistic propagator is an integral over a Lorentz scalar where the integrand is the product of a spectral weight and a free propagator containing this Lorentz scalar. One expects similar integral representation in the absence of Lorentz symmetry as well, except that the integral variable would not be scalar. An important advantage of the spectral representation is that it provides a common parametrization of causal, retarded, and advanced propagators. This feature will arise in a specially clear manner within the CTP scheme, the natural basis to introduce these different Green functions. Another issue to detail is the way the parametrization \eq{lbpar} can be carried over the spectral functions.

Let us suppose first that we have a local real field $\phi(x)$ with propagator given by Eq. \eq{spropctp}. The spectral function
\be
2\pi\rho^{(\phi)}(q)=\sum_n(2\pi)^4\delta(q-p_n)|\la0|\phi(0)|n\ra|^2
\ee
is a sum over the eigenvectors $|n\ra$ of the Hamiltonian with four-momentum $p_n^\mu$, $p^\mu_0=0$, and $p^\mu_n>0$ for $n>0$. The spectral function is vanishing for negative frequency.

We introduce the CTP matrix
\be\label{elemprop}
\hD_0(\omega,\omega')=\begin{pmatrix}\frac1{\omega^2-\omega'^2+i\epsilon}&-2\pi i\delta(\omega^2-\omega'^2)\Theta(-\omega)\cr-2\pi i\delta(\omega^2-\omega'^2)\Theta(\omega)&-\frac1{\omega^2-\omega'^2-i\epsilon}\end{pmatrix}
\ee
and elementary steps followed in the derivation of the usual spectral representation yield
\be
\hD(q)=\int_0^\infty d\omega'^2\hD_0(q^0,\omega')\rho(\omega',\v{q}).
\ee
In case of rotational invariance, the spectral strength is a two-variable function, $\rho^{(\phi)}(q)=\rho^{(\phi)}(q^0,|\v{q}|)$. Note that the spectral integral \eq{elemprop} is trivial for the CTP blocks $D^{\pm\mp}$ and the spectral function can be read off directly from the propagator since
\be
2\pi\rho^{(\phi)}(q)=iD^{-+}(q)
\ee
holds for $q^0>0$. In case of a free relativistic field of mass $m$ we find $\rho^{(\phi)}(q)=\delta(q^2-m^2)$.

The generalization of the spectral function for a local vector field $A_\mu(x)$ is a Lorentz tensor
\be
2\pi\rho_{\mu\nu}^{(A)}(q)=\sum_n(2\pi)^4\delta(q-p_n)\la0|A_\mu(0)|n\ra\la n|A_\nu(0)|0\ra
\ee
and will be characterized by the parameters
\bea\label{spectrwab}
\rho^{(A)}_\ell&=&(1-\nu^2)u^\mu u^\nu\rho_{\mu\nu},\nn
\rho^{(A)}_t&=&\hf[g^{\mu\nu}+(\nu^2-1)u^\mu u^\nu]\rho_{\mu\nu}
\eea
in case of the gauge field. This choice leads to a useful factorization of the CTP and the Lorentz indices,
\be\label{spectrwmn}
\hD^{\mu\nu}(q)=\hD_\ell(q)P^{\mu\nu}_\ell+\hD_t(q)P^{\mu\nu}_t
\ee
with
\bea\label{spectrwmnk}
\hD_\ell(q)&=&\int_0^\infty d\omega'^2\hD_0(q^0,\omega')\rho^{(A)}_\ell(\omega',\v{q}),\nn
\hD_t(q)&=&\int_0^\infty d\omega'^2\hD_0(q^0,\omega')\rho^{(A)}_t(\omega',\v{q}).
\eea
The spectral functions can again be obtained from the Wightman function, 
\be\label{spectrwmp}
iD^{-+}=2\pi\rho^{(A)}_\ell P_\ell+2\pi\rho^{(A)}_tP_t
\ee
for $q^0>0$.

\section{Electron-proton collision}\label{scattres}
The calculation of the collision process in the double time scheme gives not only expectation values and transition probabilities, but provides information about consistency and irreversibility, necessary conditions for classical limit.

\subsection{Cross section}
Once the transition probability, displayed by Eqs. \eq{treprobde}-\eq{amplsqde} for the scattering $e(p_1)+p(p_2)\to e(q_1)+p(q_2)$, is found, then the construction of the cross section is straightforward and standard. Averaging over the initial and sum of the final fermion polarizations in Eqs. \eq{treprobde}-\eq{amplsqde} yields
\bea\label{trprobf}
|{\cal T}_d|^2&=&\frac{e^4}{4m^2M^2}\bigl\{2\Re[(p_1D^{++}(r)p_2)(q_1D^{++*}(r)q_2)]+2\Re[(p_1D^{++}(r)q_2)(q_1D^{++*}(r)p_2)]\nn
&&+\tr[D^{++}(r)D^{++*}(r)](m^2-p_1q_1)(M^2-p_2q_2)\nn
&&+2(M^2-p_2q_2)\Re[p_1D^{++}(r)D^{++*tr}(r)q_1]+2(m^2-p_1q_1)\Re[p_2D^{++tr}(r)D^{++*}(r)q_2]\bigr\},\nn
|{\cal T}_e|^2&=&\frac{e^4}{4m^2M^2}\bigl\{2\Re[(p_1D^{+-}(r)p_2)(q_1D^{+-*}(r)q_2)]+2\Re[(p_1D^{+-}(r)q_2)(q_1D^{+-*}(r)p_2)]\nn
&&+\tr[D^{+-}(r)D^{+-\dagger}(r)](m^2-p_1q_1)(M^2-p_2q_2)\nn
&&+2(M^2-p_2q_2)\Re[p_1D^{+-}(r)D^{+-\dagger}(r)q_1]+2(m^2-p_1q_1)\Re[q_2D^{-+\dagger}(-r)D^{-+}(-r)p_2]\bigr\},
\eea
where $r=q_1-p_1$. The cross section is finally obtained in the usual fashion, 
\be\label{crosss}
\sigma=\frac{I}{4\sqrt{(p_1\cdot p_2)^2-m^2M^2}}
\ee
where the intensity of the transition,
\be
I=\int_S\frac{d^3q_1d^3q_2mM}{(2\pi)^6\omega(q_1)\Omega(q_2)}(2\pi)^4\delta(q_1-p_1+q_2-p_2)|{\cal T}|^2
\ee
with $\omega(q)$ given after Eq. \eq{xvac} and $\Omega(q)=\sqrt{M^2+q^2}$. The integration is over the region $S$ in the final momentum space covered by the detector.

The cross section \eq{crosss} is manifestly Lorentz invariant in the vacuum, $\kfer=0$. In the presence of environment, the electron gas, it depends on the average momentum of the environment. When the gas is at rest in the center-of-mass frame of the colliding particles, then the exchanged photon has vanishing energy. This makes the entanglement contributions to the cross scattering, $|{\cal T}_e|^2$, vanishing. 

To see the effects of entanglement, we choose both the colliding electron and the electron gas at rest in the laboratory frame. In other words, we consider an experiment where an electron gas is taken as target into a proton beam and those scattering events are filtered out where the initial electron is in the state $\v{p}=0$. The differential cross section in this frame, corresponding to a given beam particle momentum $\v{p}_2$ and scattering angle $d\Omega$, is
\be\label{diffcrsr}
\frac{d\sigma^2}{d|\v{p}_2|d\Omega}=\frac{M|\v{q}_1|}{16\pi^2|\v{p}_2|(m+\Omega(\v{p}_2)-\omega(\v{q}_1))}|{\cal T}|^2.
\ee
Cross sections, calculated by using the transition probability $|{\cal T}_d|^2$ or $|{\cal T}_e|^2$ in this expression will be denoted by $\sigma_d$ and $\sigma_e$, respectively.

The natural parameters of the cross section \eq{diffcrsr} are the beam momentum and the scattering angle. But the kinematical regime allowed on this two-dimensional plane is not a simple rectangle. To simplify the independent variables, we bring the system into the center-of-mass frame for the colliding charges by performing an appropriate Lorentz transformation. Since the coefficient of the transition probability on the right-hand side of Eq. \eq{diffcrsr} is Lorentz invariant, all we need to do is to parametrize the scattering amplitude by the natural variables of the center-of-mass frame, namely, the charged particle momentum $P$ and scattering angle $\Theta$. The particle momenta in the center-of-mass system, written as
\bea
P_1&=&(\Omega(P),-P,0,0),~~~Q_1=(\Omega(P),-P\cos\Theta,-P\sin\Theta,0),\nn
P_2&=&(\omega(P),P,0,0),~~~~~Q_2=(\omega(P),P\cos\Theta,P\sin\Theta,0),
\eea
is used to express momenta in the rest frame of the electrons,
\bea
p^x&=&\frac{P^x-vP^0}{\sqrt{1-v^2}},\nn
p^0&=&\frac{P^0-vP^x}{\sqrt{1-v^2}},
\eea
where the boost velocity in the $x$-direction is $v=P/\omega(P)$. We arrive in this manner at the expressions
\bea
p_1^x&=&-\frac{P+v\Omega_P}{\sqrt{1-v^2}},\nn
\v{q}_2&=&\left(\frac{P\cos\Theta-v\Omega_P}{\sqrt{1-v^2}},P\sin\Theta\right).
\eea
The cross section \eq{diffcrsr} now reads as
\be\label{diffcrs}
\frac{d\sigma^2}{dPd\Omega}=\frac{|\v{q}_1|(\Omega(P)+\omega(P))}{16\pi^2|\v{p}_2|(m+\Omega(\v{p}_2)-\omega(\v{q}_1))}\left[1+\frac{P^2}{\Omega(P)\omega(P)}\right]|{\cal T}|^2.
\ee

\begin{figure}
\parbox{5.5cm}{\includegraphics[scale=.4]{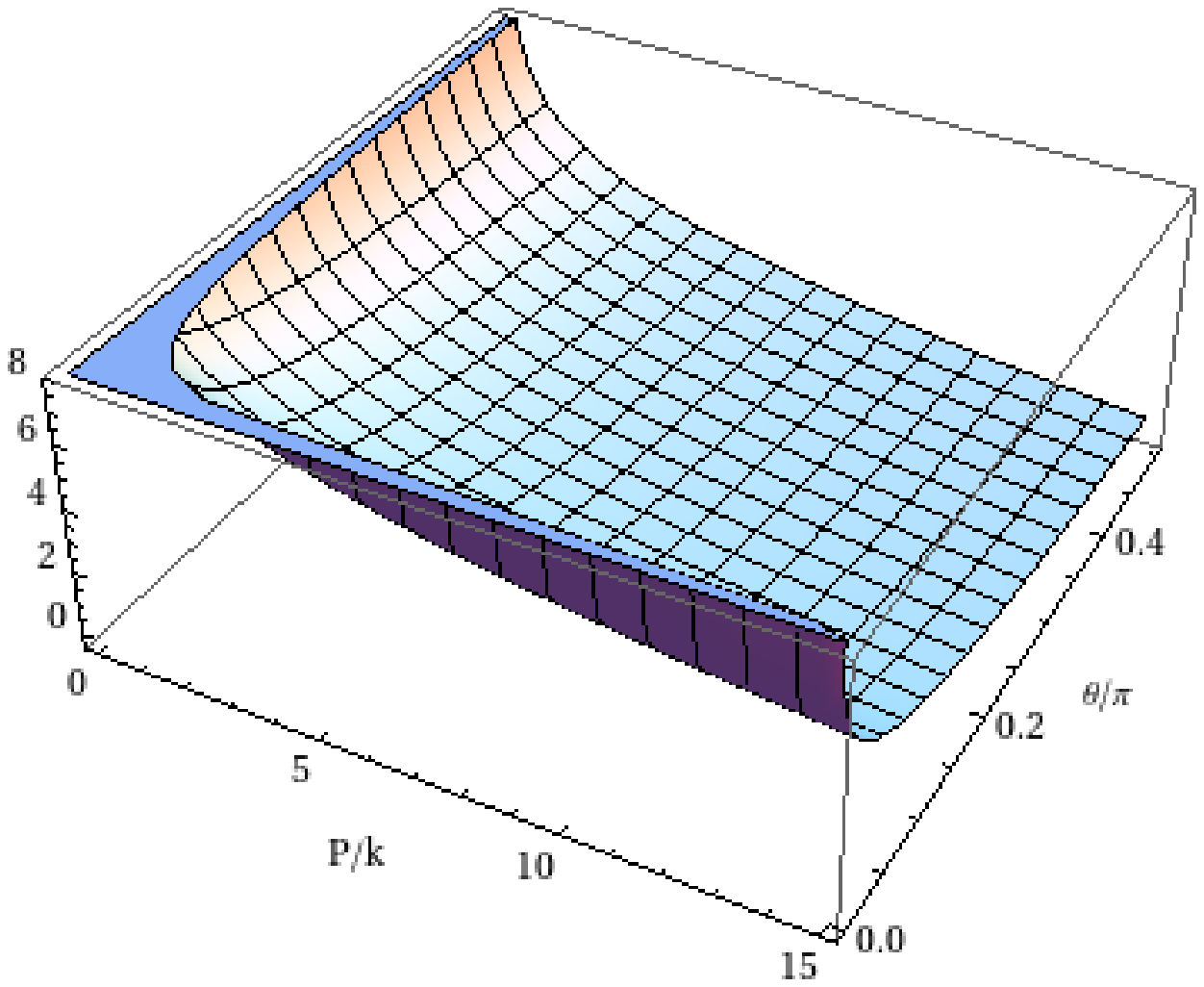}\\(a)}
\parbox{5.5cm}{\includegraphics[scale=.4]{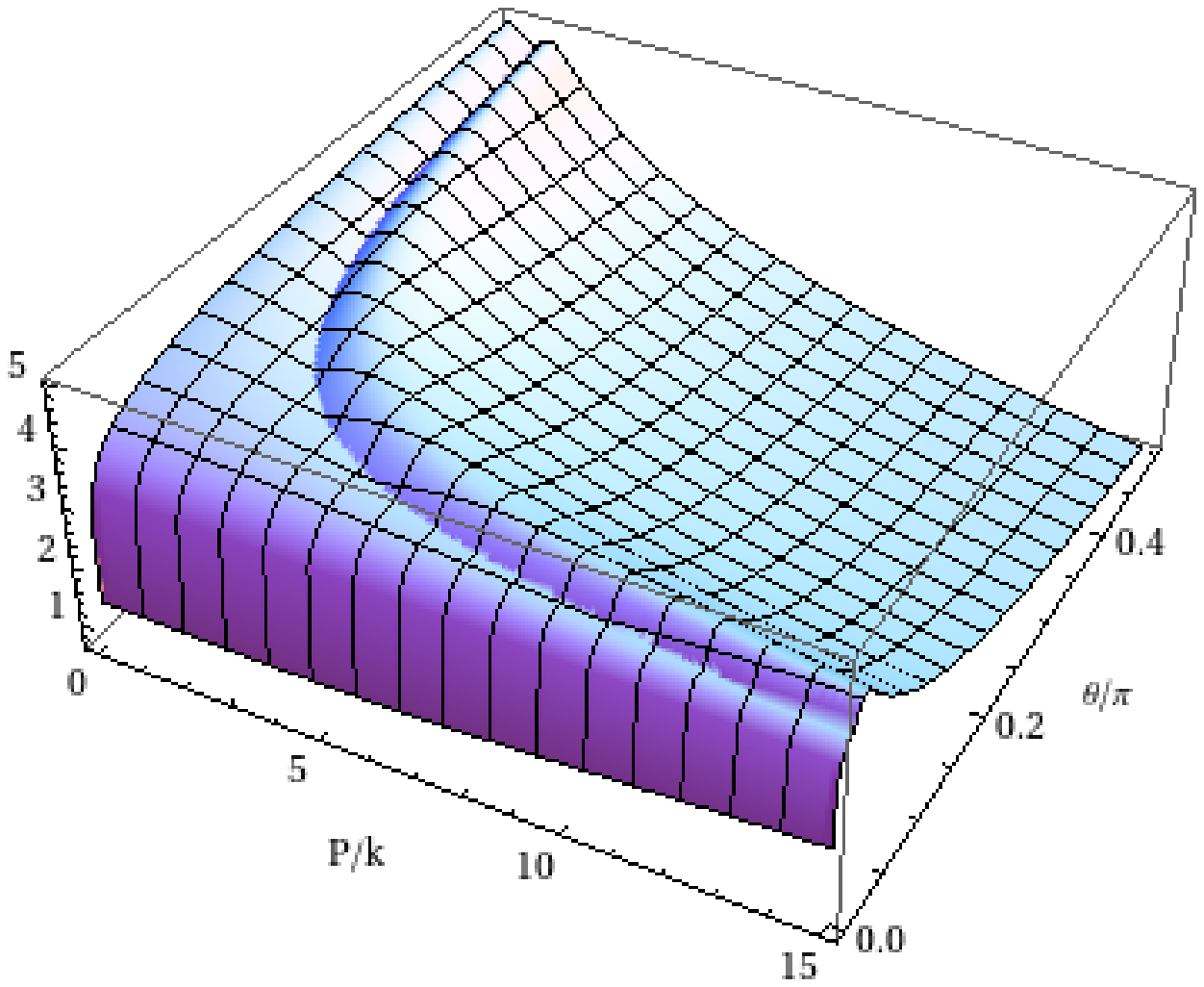}\\(b)}
\parbox{5.5cm}{\includegraphics[scale=.4]{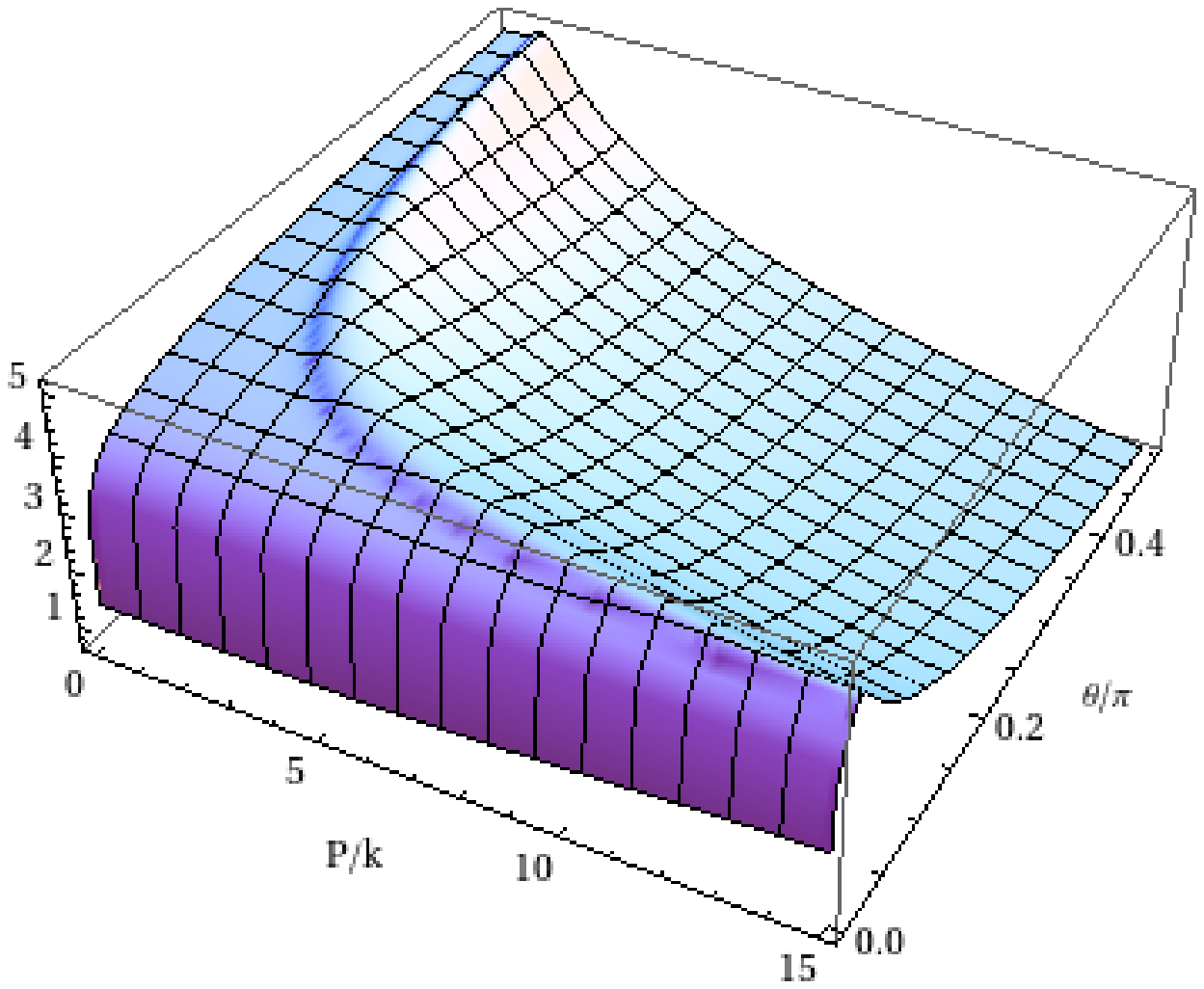}\\(c)}
\caption{Logarithm of the differential cross sections (a) $\log\sigma_v$, (b) $\log\sigma_d$ and (c) $\log(\sigma_d+\sigma_e)$ as functions of $P/\kfer$ and $\Theta/\pi$.}\label{svsn}
\end{figure}

\begin{figure}
\parbox{5.5cm}{\includegraphics[scale=.4]{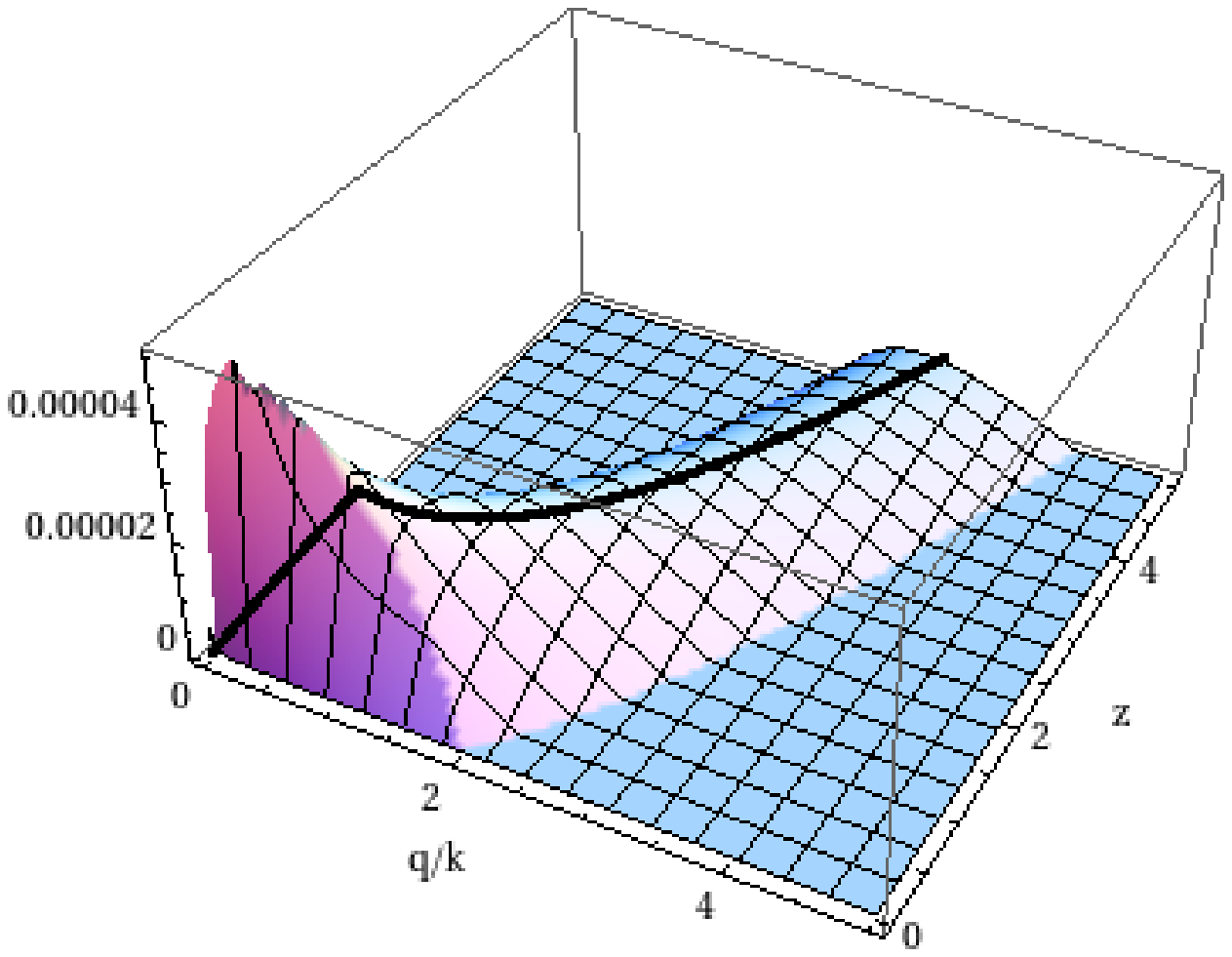}\\(a)}
\parbox{5.5cm}{\includegraphics[scale=.4]{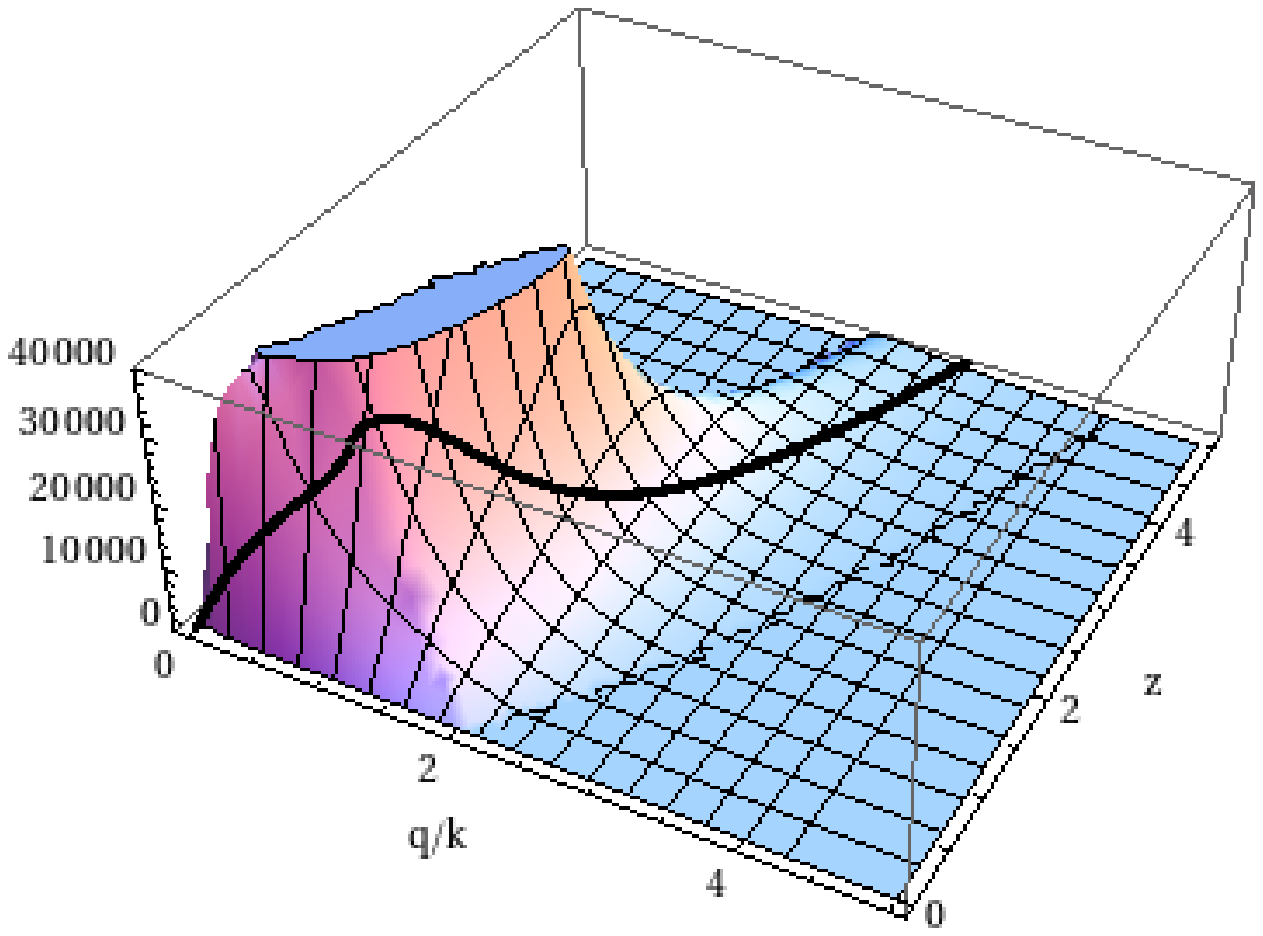}\\(b)}
\parbox{5.5cm}{\includegraphics[scale=.4]{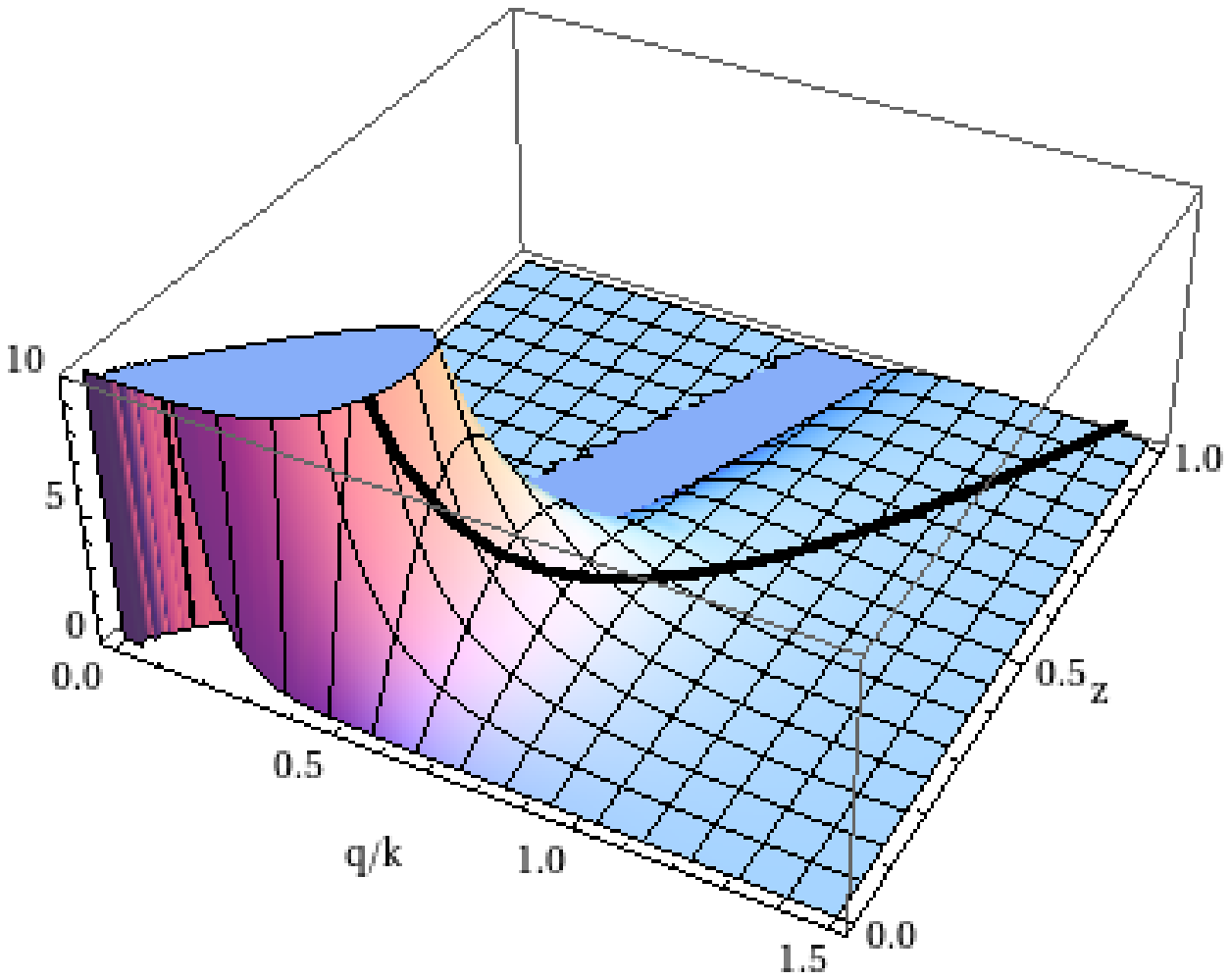}\\(c)}
\caption{Spectral strengths (a) $2\pi\rho^{(j)}_\ell$ of the electric current, (b) $2\pi\rho^{(A)}_\ell$ and (c) $2\pi\rho^{(A)}_t$ of the photon propagator as functions of $|\v{q}|/\kfer$ and $z=mq^0/\kfer^2$. The additional curve belongs to $\Theta=0.5\pi$.}\label{dpm}
\end{figure}

\begin{figure}
\parbox{5.5cm}{\includegraphics[scale=.4]{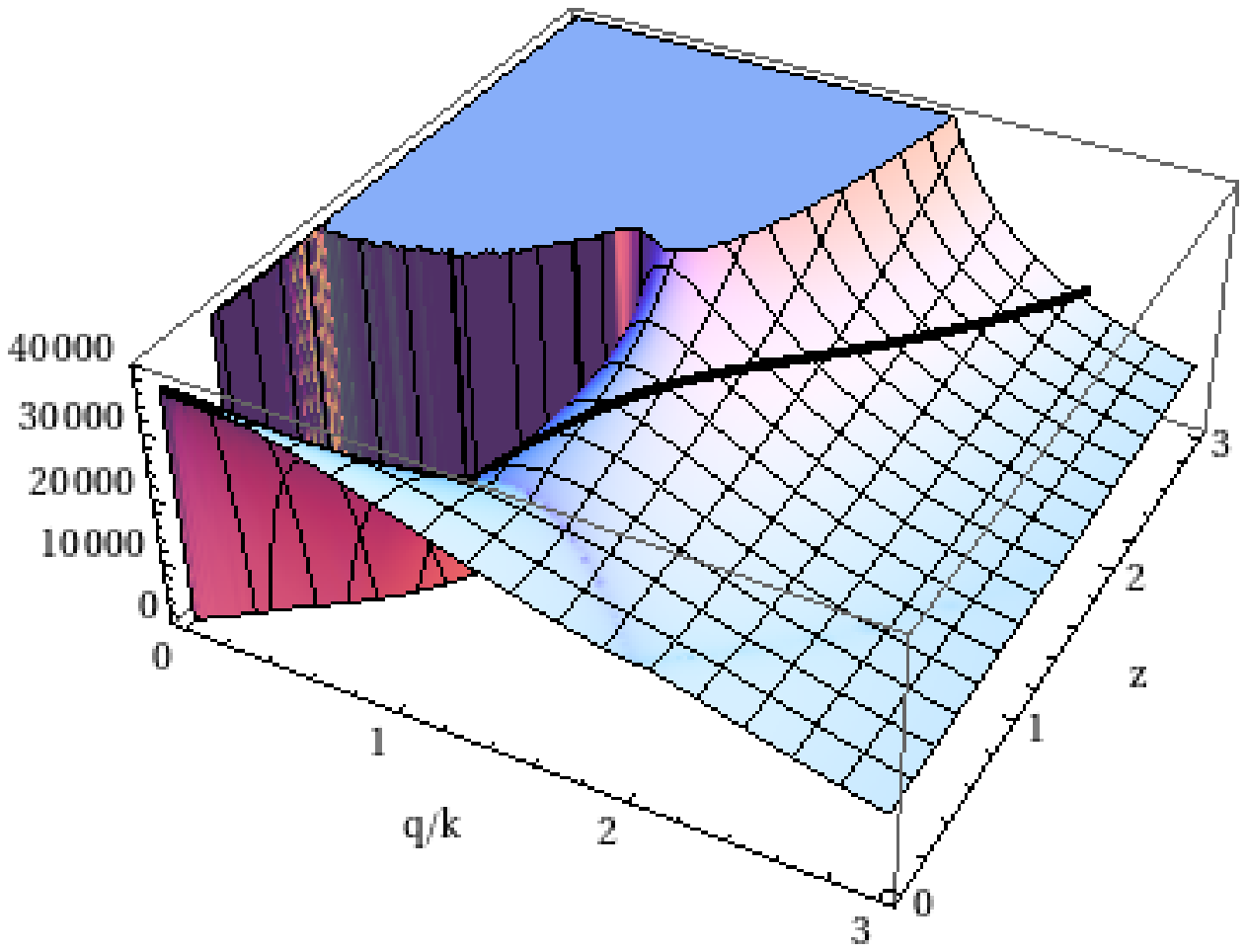}\\(a)}
\parbox{5.5cm}{\includegraphics[scale=.4]{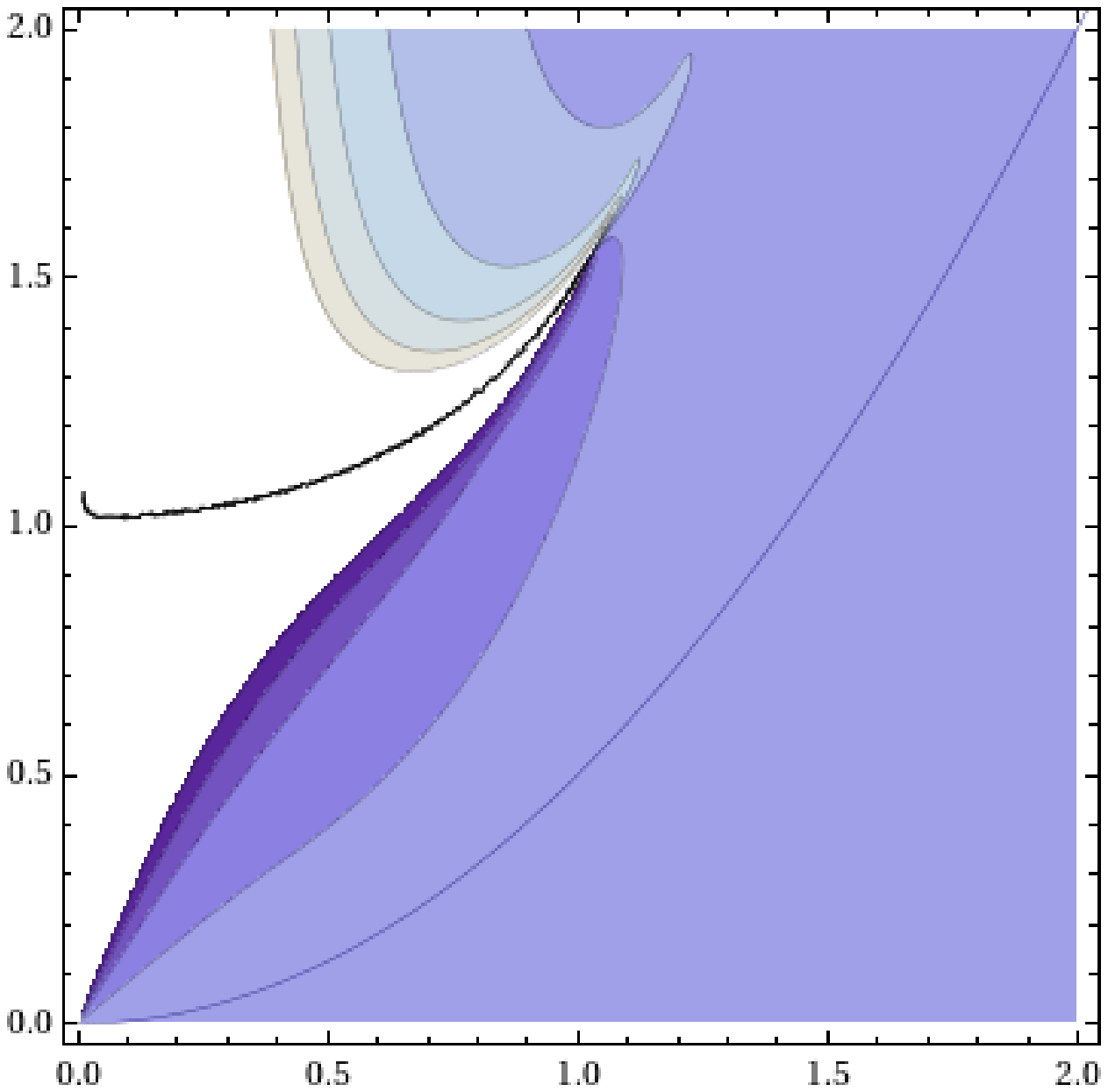}\\(b)}
\parbox{5.5cm}{\includegraphics[scale=.4]{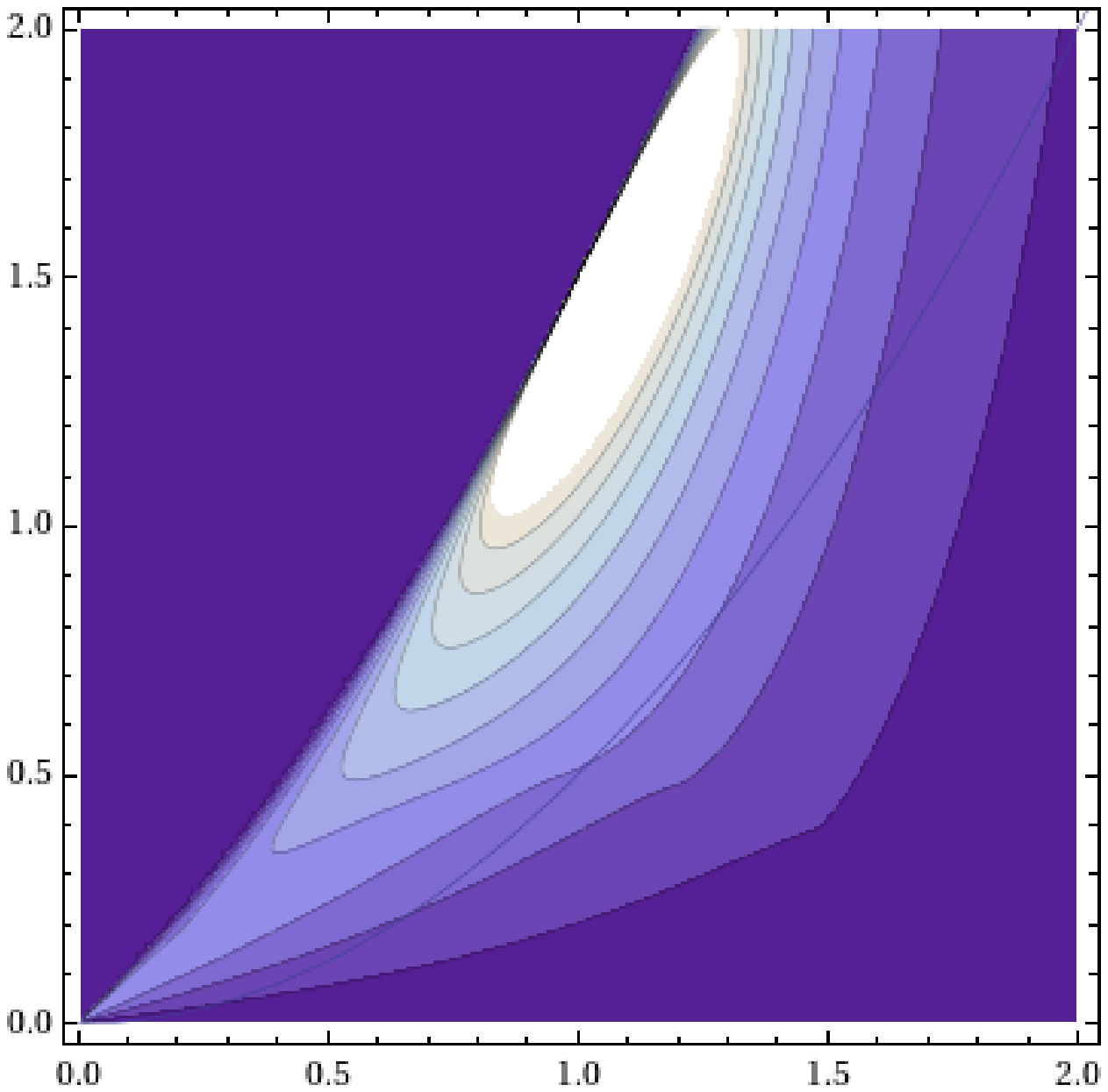}\\(c)}
\caption{(a) $D^n_\ell$ and contourplots of (b) $D^n_\ell$, (c): $2\pi\rho^{(A)}_\ell$ as functions of $|\v{q}|/\kfer$ and $z=mq^0/\kfer^2$. The line $\Theta=0.5\pi$ is shown, as well.}\label{contourpl}
\end{figure}

\begin{figure}
\parbox{5cm}{\includegraphics[scale=.4]{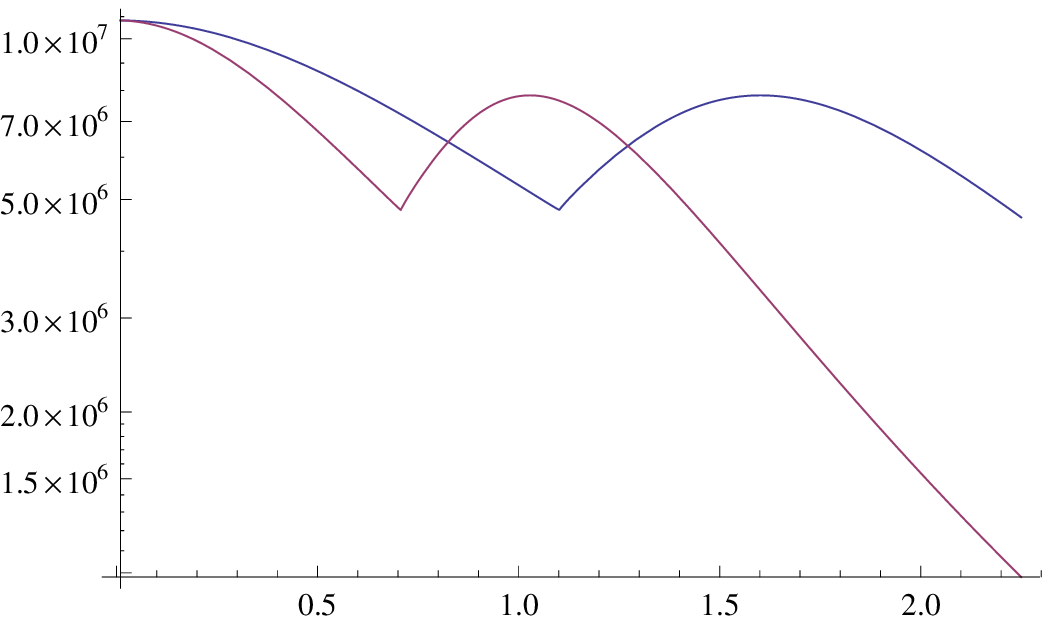}\\(a)}
\parbox{5cm}{\includegraphics[scale=.4]{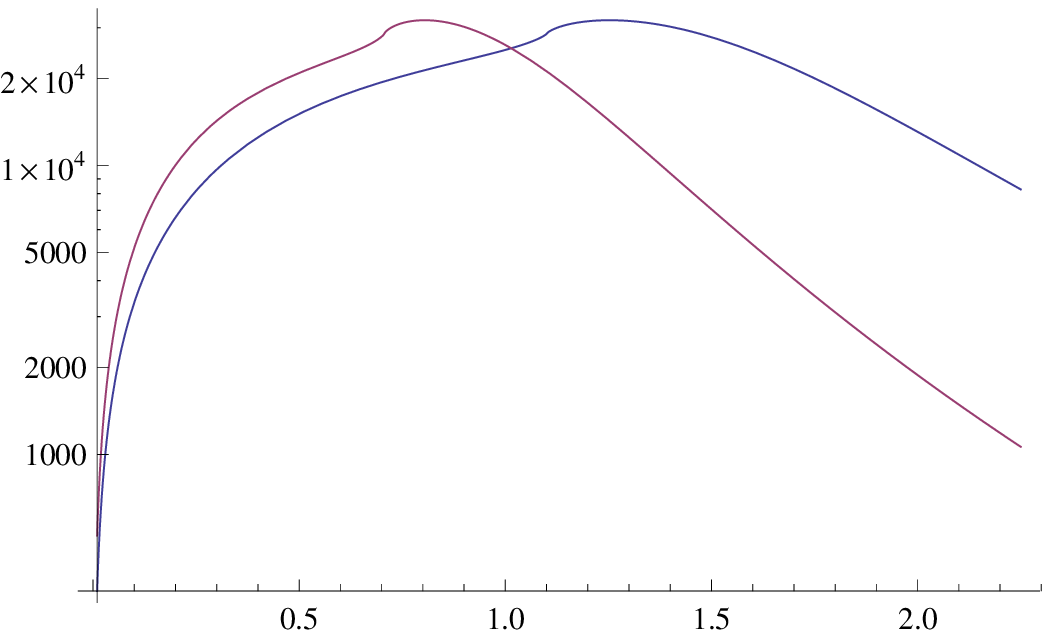}\\(b)}

\parbox{5cm}{\includegraphics[scale=.4]{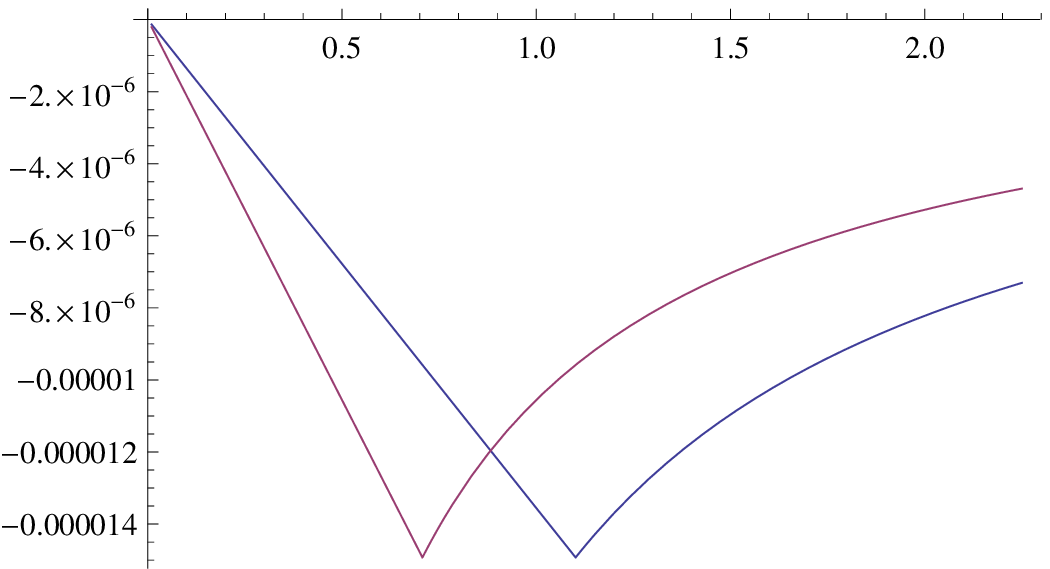}\\(c)}
\parbox{5cm}{\includegraphics[scale=.4]{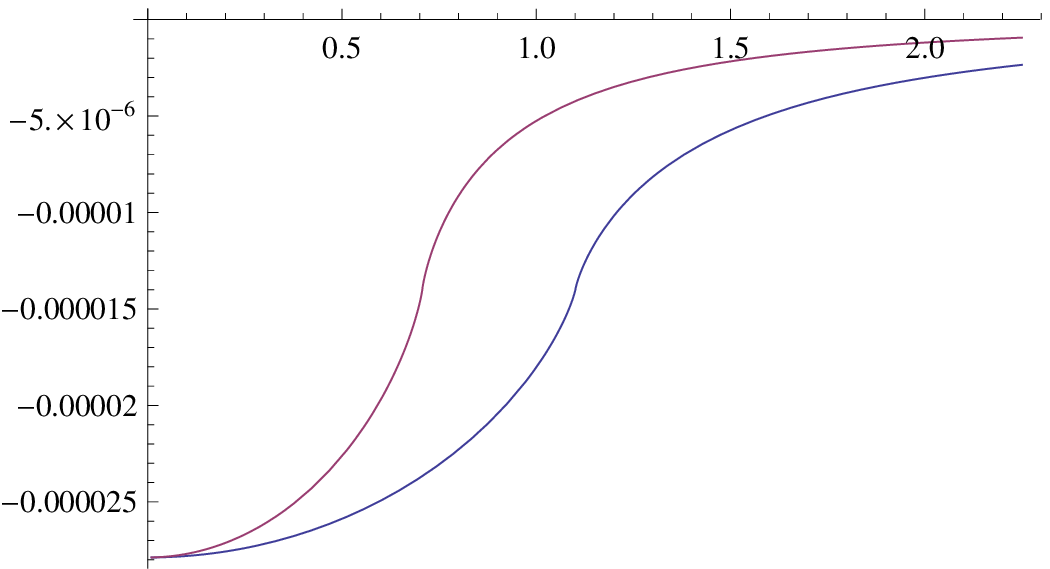}\\(d)}

\parbox{5cm}{\includegraphics[scale=.4]{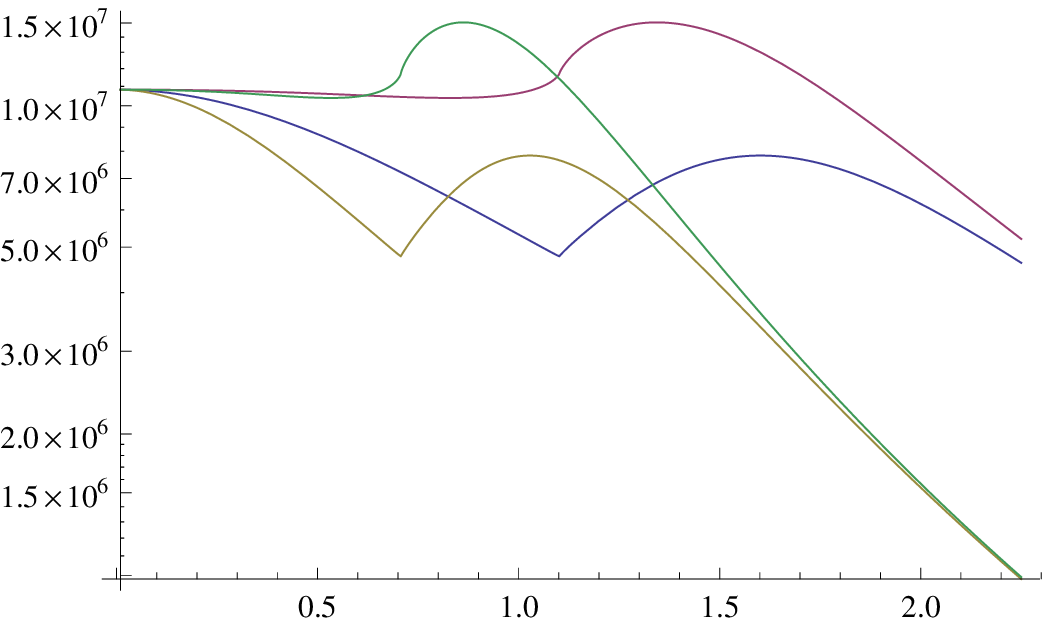}\\(e)}
\parbox{5cm}{\includegraphics[scale=.4]{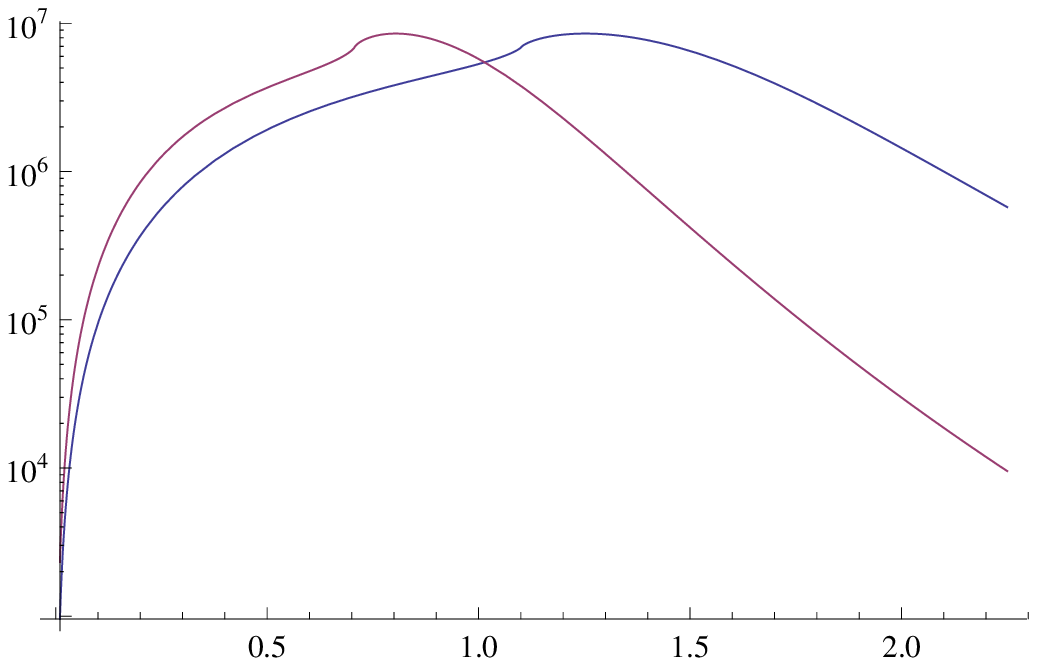}\\(f)}
\caption{(a) $|{\cal T}_d|^2$, (b) $2\pi\rho^{(A)}_\ell$, (c) $\Pi^n_\ell$, (d) $2\pi\rho^{(j)}_\ell$, (e) $|{\cal T}_d|^2+|{\cal T}_e|^2$ (higher curves), $|{\cal T}_d|^2$ (lower curves), and (f) $|{\cal T}_e|^2$ as functions of $P$ at $\Theta=0.3\pi$ (left curve) and $\theta=0.5\pi$ (right curve).}\label{sivpl}
\end{figure}

\subsection{Diagonal contribution}
The numerical results, presented below have been obtained for $\kfer=0.003m$, corresponding to $2.25$eV Fermi energy, typical metallic density at the lower edge of applicability of perturbation expansion. 

The differential cross section $\sigma_v$ in the vacuum, $\kfer=0$, is shown in Fig. \ref{svsn} (a) as the function of dimensionless center-of-mass momentum $P/\kfer$ and scattering angle $\Theta/\pi$. It diverges when $P\to0$ or $\Theta\to0$, but its value is cut off in the plot for clarity. The cross section, $\sigma_d$, depicted in Fig. \ref{svsn} (b), shows two obvious effects. One is screening, a suppression for small energy exchanges. The other is a valley and a rim appearing at a certain exchange energy. 

It is instructive at this point to look into the spectral strengths. The spectral strength $2\pi\rho^{(j)}_\ell$, introduced in Eq. \eq{spectrwab} is a measure of the amount of particle-hole states available for the longitudinal electric current operator. It is shown in Fig. \ref{dpm} (a) as the function of the dimensionless momentum $|\v{q}|/\kfer$ and frequency $z=mq^0/\kfer^2$. It is nonvanishing around the free-particle dispersion relation and displays a steep, linear increase at low frequencies within the momentum range $|v{q}|<2\kfer$ ending at a sharp edge. This and the transverse spectral strength of the current are built into the photon propagator in a nonlinear manner, as indicated by Eq. \eq{retadvpr}. The resulting spectral strengths for the photon propagator are given in Figs. \ref{dpm} (b) and (c). 

Another curve is shown in the figures, as well; it is the projection of the line parametrized by the momentum $P$ in the center-of-mass system at $\Theta=0.5\pi$. The nonlinearity of the Schwinger-Dyson resummation transforms and shifts the maximum of Fig. \ref{dpm} (a) towards larger frequencies in the photon spectral strength seen in Fig. \ref{dpm} (b). The curve $\Theta=0.5\pi$ reaches its maximum at $|\v{q}|/\kfer\sim1$, at the local maximum of $\sigma_d$ in Fig. \ref{svsn} (b) at the same scattering angle. The spectral weight of transverse photons, shown in Fig. \ref{dpm} (c), displays a narrow diverging peak at the origin and the line $\Theta=0.5\pi$ slides down on it as $P$ increases. This singularity is suppressed by the momentum variables multiplying the propagator and gives small contribution around the rim in Fig. \ref{svsn} (b), which is due to the increased number of particle-hole states to mix with longitudinal photons.

The electric current spectral strength display maximum around quasiparticles, collective particle-hole excitations, defined by the root of $\Re(D^{++})^{-1}$ on the energy-momentum plane. The function $D^n_\ell$ and its contour plot are shown in Figs. \ref{contourpl} (a) and (b) together with the line $\Theta=0.5\pi$. The propagator is divergent along the heavy line starting approximately horizontally at $z\sim1$, corresponding to plasmonlike collective excitations \cite{maxwell}. The imaginary part of the one-loop propagator is vanishing in this region; these collective modes decay slowly into several particle-hole pairs. $\Im\Pi^i_\ell$ starts to become nonvanishing when we arrive at the end of the solid line at $z\sim1.5$. The lowest line separating different colors starting from this point on the contour plot runs approximately along the line $\Re(D^{++})^{-1}=0$ corresponding to the zero-sound excitations. The longitudinal propagator, $D^n_\ell$, is finite along this quasi particle line where $\Im\Pi^i_\ell\ne0$ making the zero-sound strongly damped by single particle-hole pairs. The peak of the cross section at $P\sim1$ appears at $|\v{q}|/\kfer\sim1.4$, approximately the closest point of the end of the singular solid line to the curve $\Theta=0.5\pi$. The contour plot of $\rho^{(A)}_\ell$, shown in Fig. \ref{contourpl} (c), indicates that the number of available longitudinal photon states reaches a maximum at the end of the solid line of Fig. \ref{contourpl} (a), where the plasmon and zero-sound lines meet. The lesson of these results is that quasiparticles around the Fermi surface, $|\v{q}|\sim\kfer$, $z\sim1$, are responsible for the rim in the direct cross section $\sigma_d$.

To better locate the rim and valley of $\sigma_d$ in Fig. \ref{svsn} (b), two lines, one with $\theta=0.3\pi$ and another with $\theta=0.5\pi$, are followed in Fig. \ref{sivpl} as functions of the center-of-mass momentum $P$ in different terms of the transition probability. It is better to look into the transition probability rather than the cross section because the former has no kinematical factors besides the expectation value \eq{trprobop}. The transition probabilities at $\theta=0.3\pi$ and $\theta=0.5\pi$ are shown in Fig. \ref{sivpl} (a). Their peak is approximately at the maximum of $\rho^{(A)}_\ell$, shown in Fig. \ref{sivpl} (b), supporting the remark made before about the coincidence of the maximum of the cross section and of the number of longitudinal photon states. The valley of the cross section seem to agree with the position of the maximum of $\rho^{(j)}_\ell$, shown in Fig. \ref{sivpl} (c). This is rather natural; faster-decaying particle-hole states make weaker transition probability. The sharp rim of $\rho^{(j)}_\ell$ separates two different kinematical regions, the last two cases in the definition of $N$ in \eq{lmselfen}. This makes a characteristic, continuous but nondifferentiable singularity for the transition probability. It is the same location where the self-energy $\Pi^n_\ell$, depicted in Fig. \ref{sivpl} (d), has inflection point.

\subsection{Entanglement}
We now inspect the entanglement contribution to the cross section. According to the remark about the parity of the number of $\Sigma^{\pm\mp}$ factors in $\sigma_d$ and $\sigma_e$ whenever the latter dominates the former the colliding particle-gas entanglement, encoded by $\Sigma^{\pm\mp}$, is important. The complete cross section, $\sigma_d+\sigma_e$, is displayed in Fig. \ref{svsn} (c). The entanglement contribution seems to change the cross section mainly along the rim of Fig. \ref{svsn} (b). One sees the entanglement contribution clearer in Fig. \ref{sivpl} (e) where $\sigma_d+\sigma_d$ is plotted together with $\sigma_d$. The closeness of the shape of $|{\cal T}_e|^2$ and $2\pi\rho^{(A)}_\ell$ , plotted in Figs. \ref{sivpl} (f) and (b), is remarkable.

Note that the entanglement contribution to the cross section, the second line of Eqs. \eq{amplsqde}, is positive definite in agreement of the notation, $|{\cal T}_e|^2$, used. Such a definite feature of the entanglement contribution appears natural since the asymptotic particle-holes states of the gas induced by the colliding particles should always increase the cross section.

\begin{figure}
\parbox{5.5cm}{\includegraphics[scale=.4]{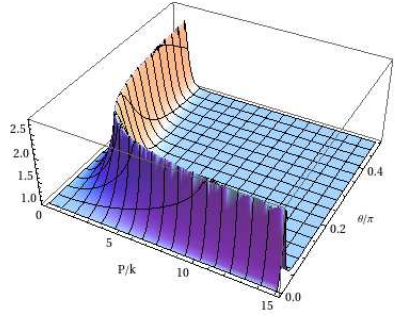}}
\caption{$(\sigma_d+\sigma_e)/\sigma_d$ as function of $P/\kfer$ and $\Theta/\pi$.}\label{snse}
\end{figure}

Finally, the ratio $(\sigma_d+\sigma_e)/\sigma_d$, depicted in Fig. \ref{snse}, shows the relative importance of the entanglement contributions around the rim of $\sigma_d$, where the particle-hole components of the exchanged photon are around the Fermi surface. The nontrivial final states of the environment, taking into account by the entanglement contribution, represent the dominant contribution to the cross section in this kinematical region.

\subsection{Classicality}
There are two known necessary conditions of quantum-classical crossover, irreversibility and decoherence. As mentioned above, the latter corresponds to a property of the state at a given time and its building up in time can be seen as consistency. 

\begin{figure}
\parbox{7cm}{\includegraphics[scale=.4]{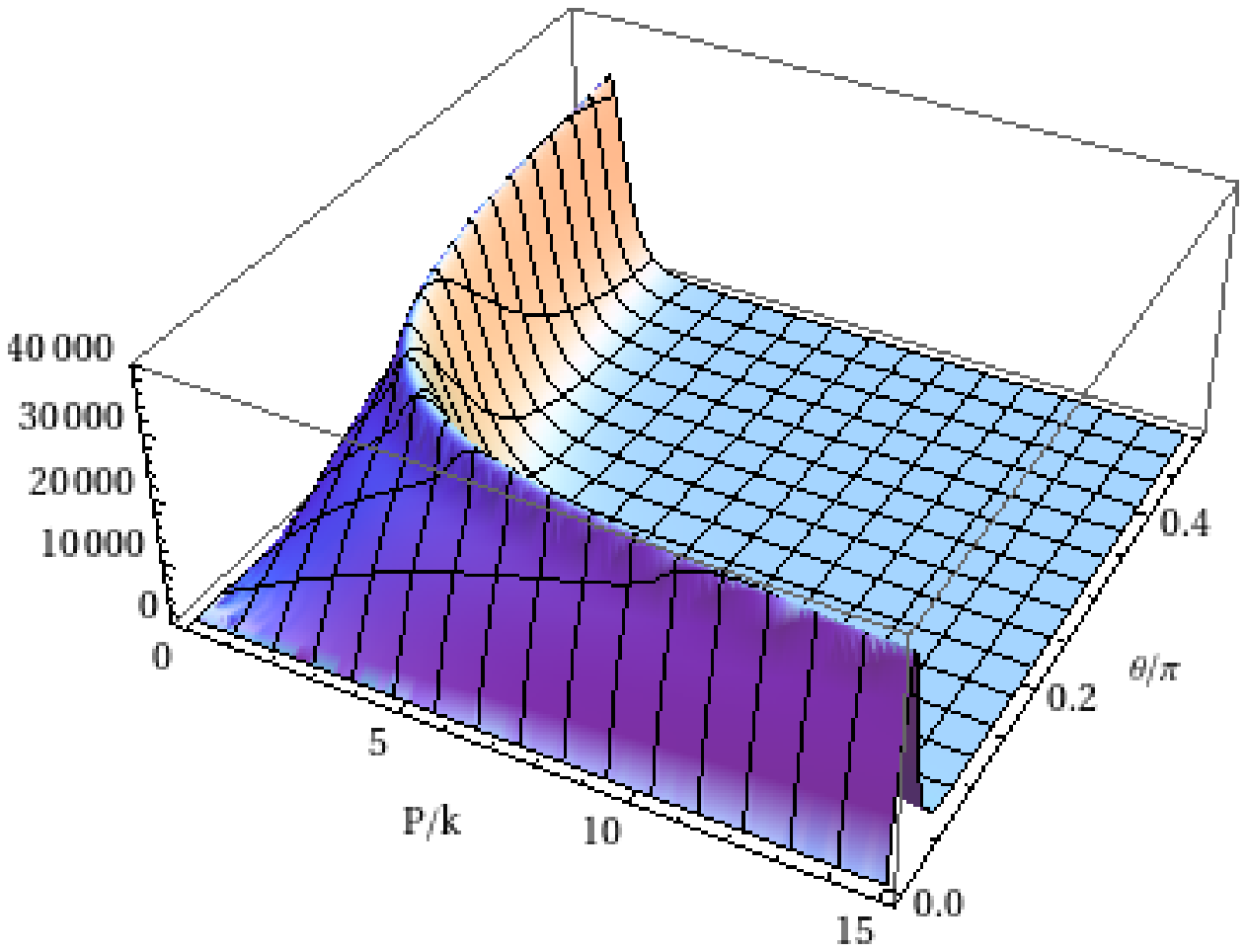}\\(a)}
\parbox{7cm}{\includegraphics[scale=.4]{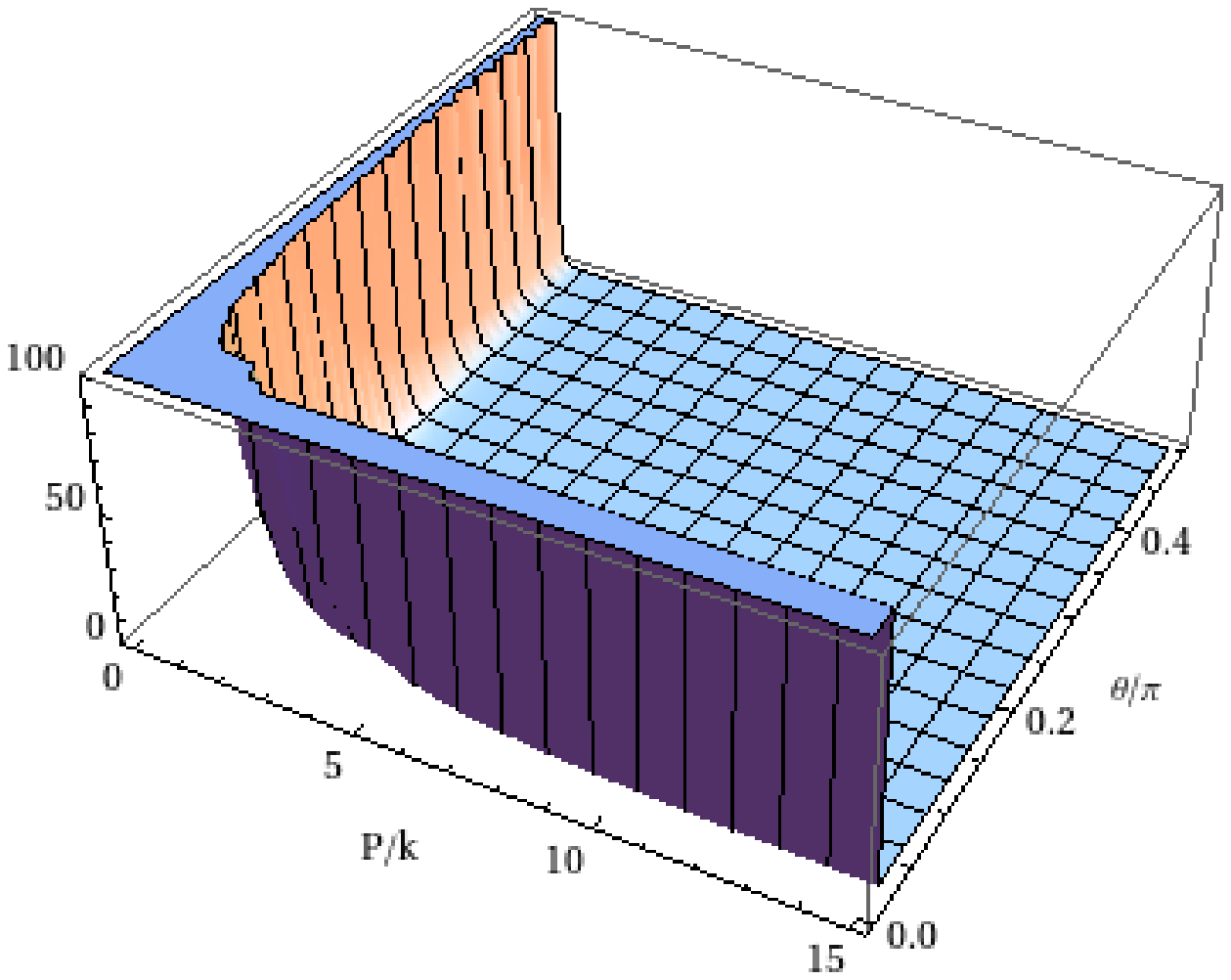}\\(b)}\hskip1cm
\caption{(a) $D^i_\ell$, (b) $D^i_t$ as functions of $P/\kfer$ and $\Theta/\pi$.}\label{cl}
\end{figure}

To simplify matter, let us consider $n$ nonrelativistic charges. The generator functional like the one defined by Eq. \eq{wpint} for the reduced density matrix of the charges is
\be
Z=\int D[\hA]D[\hat{\v{x}}]e^{\frac{i}{2}\hA\cdot\hD^{-1}_0\cdot\hA+iS[\v{x}^+]-iS[\v{x}^-]-ie\int dy\hat j[y;\v{x}]\hat\sigma\hA(y)}
\ee
where the boundary condition in time is OTP for particle trajectories $\v{x}$ and CTP for the environment, the electromagnetic field, and the electron gas. The electric current density of the given particle trajectories is denoted by $\hat j^\mu(y)=\hat j^\mu[y;\v{x}]$ and the external sources are suppressed for simplicity. Integration over the environment variables yields the effective theory \cite{ed}
\be
Z=\int D[\hat{\v{x}}]e^{iS_{eff}[\hat{\v{x}}]}
\ee
where the effective action,
\be\label{effs}
S_{eff}[\hat{\v{x}}]=S[\v{x}^+]-S[\v{x}^-]+W^\gamma[-ej[\v{x}^+],ej[\v{x}^-]],
\ee
contains the influence functional, the generator functional of the electromagnetic field,
\be
W^\gamma[\hj]=-\frac{e^2}2\int dxdy\hj^\mu(x)\hD_{\mu\nu}(x,y)\hj^\nu(y)+\ord{\hj^3}.
\ee
The form \eq{spropctp} for the photon propagator yields for the quadratic part
\bea
\Re W^{\gamma(2)}[\hj]&=&-\frac{e^2}2\int dxdy\left[j^\mu(x)D^a_{\mu\nu}(x,y)\bar j^\nu(y)-\bar j^\mu(x)D^r_{\mu\nu}(x,y)j^\nu(y)\right]\nn
\Im W^{\gamma(2)}[\hj]&=&-2e^2\int dxdy\bar j^\mu(x)D^i_{\mu\nu}(x,y)\bar j^\nu(y)
\eea
where the parametrization $j^\pm=j/2\pm\bar j$ is used. It is remarkable that the imaginary part is positive semidefinite, $\Im W^{\gamma(2)}[\hj]\ge0$. The actual form is
\be
\Im W^\gamma[-ej[\v{x}^+],ej[\v{x}^-]]=-2e^2\int\frac{dq}{(2\pi)^4}\left[\frac{D^i_\ell(q)}{\nu^2-1}|j^0(q)-\nu\v{n}\v{j}(q)|^2+D^i_t(q)(|\v{j}(q)|^2-|\v{j}(q)\v{n}|^2)\right],
\ee
where the current in the right-hand side is $j^\mu(y)=j[y;\v{x}^+]-j[y;\v{x}^-]$. 

The functions $D^i_\ell$ and $D^i_t$ are shown in Figs. \ref{cl}. The sharp rim of $D^i_\ell$ at the same position as in Fig. \ref{snse} indicates that the quasiparticles contributing strongly to entanglement scattering make the colliding particle trajectories consistent. The longitudinal, Coulomb-like contribution to consistency, shown in Fig. \ref{cl} (a), is weak at small energy-momentum transfer due to screening. In this regime the transverse radiation field starts to generate consistency according to Fig. \ref{cl} (b). One expects an enhancement of these effects by further radiative corrections not considered in this work, including soft photons in the final state.

Irreversibility is generated by the finite life time of quasi particles which is inversely proportional to the imaginary part of the inverse of the Feynman propagator at the quasiparticle dispersion relation. The CTP structure \eq{spropctp} of the photon propagator assures that both consistency and irreversibility are governed by the same dynamics, comprised in $D^i$. The main lesson is that scattering processes with large values of $D^i$, having fast-decaying quasiparticles and thereby strong irreversibility, tend to be classical.

\section{Summary}\label{summary}
A collision process with open environment channels in the asymptotic out-state sector, in particular, electron-proton collision in a nondegenerate electron gas, is treated in this work within the framework of the CTP formalism. The transition probability is obtained by the simple repetition of steps, followed in the reduction formulas and the environment is taken into account by using Schwinger-Dyson resummed photon propagator containing the one-loop self-energy. The dynamics of asymptotic environment states arises from the back-reaction of the collision process on its environment and can easily be handled by the algebraic solution of the Schwinger-Dyson-Kadanoff-Baym equation. 

It is found that back-reaction is important and dominates the cross section when the exchanged energy is around the Fermi level. Asymptotic environment states and back-reaction yield strong colliding particle-environment entanglement in this regime. Hence, whenever soft back-reaction is dominant the collision is closer to being classical and is irreversible.

One may wonder how further partial resummation of the perturbation series changes the results. Higher-order terms in the photon self-energy include multiple particle-hole pairs and, therefore, should enhance entanglement and classical features. It remains to be seen how vertex corrections modify the results. It seems reasonable to expect that our results remain qualitatively similar for degenerate electron gas and finite temperature effects should further strengthen entanglement and classical behavior.

We believe that the results make the reevaluation of cross sections necessary for elementary processes in laboratory and in extraterrestrial plasma if their kinematical regime is close to the scale of the environment. Furthermore, this method may ultimately lead to an improved description of collision processes where the multiparticle aspects of the beam and the target can be taken into account and the construction of more powerful phenomenological kinetic models to describe the quantum-classical crossover.

\end{document}